\documentclass[preprintnumbers,prd,floatfix,superscriptaddress,nofootinbib,notitlepage]{revtex4-1}
\pdfoutput=1
\usepackage{graphicx,epsfig}
\usepackage{amsmath}
\usepackage {amssymb}

\usepackage{subcaption}
\usepackage{multirow}
\usepackage{dcolumn}
\usepackage{bm}
\usepackage{amsfonts}
\usepackage{soul}

\usepackage{color}
\usepackage{relsize}

\begin{document}


\title{Recursive Penrose
processes in electrically charged black hole spacetimes:
Backreaction and energy extraction}

\author{Duarte Feiteira}
\email{duartefeiteira@tecnico.ulisboa.pt}
\affiliation{
Department of Physics and Helsinki Institute of Physics,
Gustaf H\"allstr\"omin katu 2a, FI-00014 Helsinki, Finland \&\\
Centro de Astrof\'{\i}sica e Gravita\c c\~ao  - CENTRA,
Departamento de F\'{\i}sica,\\ Instituto Superior T\'ecnico - IST,
Universidade de Lisboa - UL,\\ Av. Rovisco Pais 1, 1049-001
Lisboa, Portugal}

\author{Jos\'e P. S. Lemos}
\email{joselemos@ist.utl.pt}
\affiliation{Centro de Astrof\'{\i}sica e Gravita\c c\~ao  - CENTRA,
Departamento de F\'{\i}sica, Instituto Superior T\'ecnico - IST,
Universidade de Lisboa - UL, Av. Rovisco Pais 1, 1049-001
Lisboa, Portugal \&\\
Center for Astrophysics and Space Science,\\
School of Mathematical and Physical Sciences,\\
Macquarie University, Sydney, New South Wales 2109, Australia
}

\author{Oleg B. Zaslavskii}
\email{zaslav@ukr.net}
\affiliation{Department of Physics and Technology,
Kharkov V.~N.~Karazin National
University, 4 Svoboda Square, 61022 Kharkov, Ukraine}

\begin{abstract}

We study a recursive Penrose process
and the corresponding energy extraction
for the decay of electrically
charged particles in a Reissner-Nordstr\"om black hole spacetime with
anti-de Sitter (AdS) asymptotics, incorporating the backreaction on
the black hole's mass and electric charge. A recursive process
requires that the decay products remain confined within a finite
spatial region so that the emitted particles bounce back and undergo
further decay at some location, possibly
the same location as the initial decay. In
asymptotically AdS spacetimes, the confining property arises
naturally. Outgoing particles encounter an outer turning point and are
reflected inwards. Alternatively, one may impose a reflecting mirror
at a finite radius, but in AdS backreaction makes these two
confinement methods exactly equivalent.
Let $Q_n$ denote the black hole charge after $n$ decays, and define
the index $n_{\rm c}$ as the value for which the black hole's charge is zero,
$Q_{n_{\rm c}}=0$.
For $n_{\rm c}$ integer, which can be achieved for a specific choice of
initial conditions, the black hole's charge decreases step by
step and reaches exactly zero, at least momentarily, after a finite
number of decays, thereby terminating the recursive process.
However, the last particle emitted turns back, and encountering
zero charge, falls straight into the black hole. The final state is
thus a charged black hole
whose charge equals the sum of the original black hole charge with the
initial particle charge. 
There is a fine tuned possibility
in which the process finishes with an uncharged black hole
accompanied by a charged particle at rest precisely at the original
decay position, which is an unstable configuration.
For $n_{\rm c}$ noninteger, the black hole charge decreases and can be
made arbitrarily small, but it never reaches zero. The last physically
allowed decay occurs at $n=n_{c}^-$, where $n=n_{c}^-$ is the greatest
integer less than $n_{\rm c}$. Any further decay would invalidate the
approximations, one of the particles would carry a charge comparable
to the black hole mass, transforming the problem into an uncontrolled
two-body interaction. The would-be subsequent decay would moreover
violate cosmic censorship, preventing it entirely. Therefore the process
is terminated before any inconsistency arises.
In both the integer and noninteger cases, the system yields a finite
energy gain. Backreaction ensures
that the process has a finite
duration and extracts only a finite amount of energy. No black hole
bomb occurs, the system can at most
work as an energy factory. In short,
we show
that accounting for backreaction renders the black hole
bomb impossible.

\end{abstract}
\maketitle

\newpage

\section{Introduction}
\label{sec:intro}

\subsection{Penrose process: Generics}

The Penrose process is a remarkable mechanism proposed
for extracting energy from a black hole. Originally formulated for
rotating Kerr black holes, it relies on the existence of
an ergoregion, defined as the region containing negative-energy states
with respect to infinity, allowing particles to
emerge with more energy than the energy originally injected.
This energy is
ultimately extracted from the black hole's rotational energy. 
The process can be accomplished in the context of particle
disintegration, or decay,
where one fragment can fall into the black hole with negative
energy while the other escapes to infinity with an energy exceeding
that of the original particle. Generalizations consider
collisions, in which interacting particles inside the ergoregion can
send one product into the black hole with negative energy, enabling
another product to escape with enhanced energy.
These two classes, the class of disintegrational processes and the
class of collisional processes, give the only physical scenarios
capable of realizing the Penrose process, perhaps with some variants
that can be included in one or the other class.  Since in general
relativity, a black hole can have two charges, electric charge and
angular momentum, the extraction of energy in whatever class can
proceed via these charges.

\subsection{Electrically charged Penrose process}

We begin by focusing on the electric case, which is the primary
subject of this work.  Although astrophysically charged black holes
are generally considered unlikely, black holes exist across all
scales, and microscopic black holes may carry non-negligible
charge. Consequently, electrically charged black holes are of
considerable interest.  In the electric Penrose process, the black
hole carries an electric charge, and the particles involved are
likewise electrically charged. Negative-energy states arise due to the
black hole's electrostatic potential, leading to the formation of an
electric ergosphere and enabling energy extraction from 
an electric charged black hole. The interaction of charged particles
with the black hole's electromagnetic field renders this mechanism
physically well defined and well suited to the present setting.

In \cite{christodoulouruffini}, it was shown that up to
fifty percent of the
mass of an extremal charged 
Reissner-Nordstr\"om black hole black hole can, in principle, be converted
into extractable energy, placing electrically charged black holes
firmly on the theoretical map. 
Indeed, \cite{denardo} demonstrated that electrically charged black
holes possess ergoregions and studied the disintegration-type Penrose
process in this setting. The collisional Penrose process for
electrically charged black holes and charged particles, together with
an analysis of its efficiency, was discussed in
\cite{waghdadidch}. Another important type of collision,
the
Ba\~nados-Silk-West (BSW) mechanism, occurring both at the
ergosphere and at the event horizon
allowing for high energy production locally, was first studied in the
electrically charged
case in \cite{zalavskii1}, revealing the possibility of
enhanced energy extraction at infinity.

Subsequently, numerous works have explored electrically charged
Penrose processes involving charged black holes and charged
particles. In \cite{zaslavskii2012extremal}, the collisional Penrose
process was analyzed for extremal charged black holes, where the
collision is of the BSW type between charged particles. The escape of
superheavy and highly energetic particles from charged black holes was
investigated in \cite{nemoto}. A detailed study of ergoregions and
negative-energy states in charged black holes was presented in
\cite{zaslavskii2020}. It was further shown in \cite{zaslavskii2021}
that unbounded energies at infinity can be extracted in a
super-Penrose process in the electrically charged case.
The Penrose process for a charged black hole immersed in a uniform
magnetic field was studied in \cite{gupta2021}.

In \cite{kokubu2021},
the confinement of charged particles through a mirror
placed at some distance from the black hole
and some of its implications for the
Penrose process were examined.
Energy
extraction from extremal Reissner-Nordstr\"om black holes and their
generalizations via charged particle collisions was analyzed in
\cite{hejda}. Axially symmetric magnetized Reissner-Nordstr\"om black
holes and charged particles within the Penrose process were studied in
\cite{shaymatov2022}.
In \cite{zaslavskii2022}, through a recursive
disintegrational Penrose
process in Reissner-Nordstr\"om black hole spacetimes,
charged particles confined within a mirror 
were investigated, leading to a black hole bomb, where
unbound energies can be achieved leading
to the destruction of the mirror itself, a result
that was extended to neutral particles in the rotational case.
In \cite{paper1},
an additional confinement procedure
was added, namely,
a cosmological constant, and the conditions for
the existence of factories or
bombs in Reissner-Nordstr\"om
black hole spacetimes with anti-de Sitter (AdS) asymptotics
were identified.

The fundamental
properties of the electric Penrose process were summarized in
\cite{zaslavskii2024}. An account of Penrose and
super-Penrose energy extraction from Reissner-Nordstr\"om black holes
with a cosmological constant via the BSW mechanism was presented in
\cite{flz2025}.

\subsection{Rotational Penrose process}

We have been discussing Penrose processes in which electrically
charged particles extract energy from an electrically charged black
hole.
However, the Penrose process was originally formulated for Kerr black
holes, the rotating
black holes in general relativity, using a disintegrational
mechanism \cite{penrose_book,penrosefloyd}. In this scenario,
a particle entering the
ergoregion
decays into two fragments. One fragment,
possessing negative energy, falls through the event horizon, while the
other escapes to infinity with energy exceeding that of the original
particle. The energy gain is supplied by the black hole's
rotational charge, i.e., its angular
momentum. In this disintegrational
rotational case, the net extractable energy is
bounded from above at roughly twenty percent of the black hole mass
\cite{bardeen,wald}.

The collisional Penrose process for rotating black holes replaces
particle disintegrations with particle collisions within the ergoregion
and was studied in \cite{psk}.
The BSW mechanism, studied first in the rotating case
\cite{bsw},
 allows for arbitrarily large
center-of-mass energies and opens the possibility of enhanced energy
extraction at infinity in a Penrose process.
An interesting discussion of these processes
was presented in \cite{komissarov}. The maximum efficiency of the
collisional Penrose process was analyzed in \cite{zaslavski1607}.
Further developments and comprehensive reviews, including the role of
the BSW mechanism, can be found in
\cite{haradakimura,schnittman,hejda2109}.

The general properties of
the disintegrational
Penrose process involving neutral particles in the equatorial
plane were examined in \cite{zaslavski2307}.
A recursive 
disintegrational Penrose
process in Kerr black hole spacetimes, where
orbiting particles
 confined within a mirror
could lead to a black hole bomb can also be performed.
The repetitive 
Penrose process, a process in which the Penrose process
is applied concomitantly or quasi-concomitantly
for a bunch of particles, was investigated in
\cite{repetitiverpqz}.

Energy extraction can also proceed from black holes that possess
both electric charge and angular momentum, allowing the Penrose
process to tap simultaneously into both reservoirs.

\subsection{Superradiance process}

Another mechanism for extracting energy from a black hole relies not
on particles but on wave fields. In this process, incident waves are
sent into the ergoregion and emerge with increased amplitude through
superradiant scattering.
For electrically charged black holes, electric superradiance occurs
when the frequency of the incoming wave is smaller than the product of
the wave's electric charge and the electric potential of the black
hole. Under this condition, the wave is amplified, enabling energy
extraction \cite{bekenstein}.
If one further encloses the black hole within a mirror
or inside an AdS spacetime, superradiance 
can operate either as an energy factory or, in
more extreme configurations, as a black hole bomb, see, e.g.,
\cite{dhr,dolanpw,sanchis,bosch}. It has also been shown
that the endpoint of this process superradiating
may be a hairy black hole
\cite{yanghuang}.
For rotating black holes, superradiance including  the
possibility of a bomb was studied in
\cite{teukpress,starob,cdly}. In this case, amplification occurs when the
wave frequency is smaller than the product of its azimuthal quantum
number and the angular velocity of the black hole,
the extracted energy comes at the cost of the
black hole's rotational energy.

\subsection{Recursive processes}

In both the electrically charged and rotating cases, it is possible to
engineer a recursive process that enhances the extracted energy,
as some of the works cited above show. If a
reflective mirror surrounds the black hole, a particle or wave
propagating outward from the ergoregion is reflected back toward it,
where it can undergo amplification. Recursive reflections lead to
successive amplification cycles, causing the extracted energy to grow
geometrically. Two outcomes are then possible. Either the
black hole exhausts its available energy reservoir
leading to an energy factory, or
the mounting
pressure destroys the mirror in an explosive event leading
to the black hole bomb.
There are other mechanisms that
can effectively replace the reflective mirror. One
possibility, that works both
in the Penrose process and superradiance, is to consider
a spacetime with a negative
cosmological constant. In this case,
the negative cosmological constant supplies
an intrinsic confining structure, replacing the need for an
external reflective mirror. Thus, black holes in
asymptotically AdS
spacetimes are of interest in this context.
Another possibility that  also works both
in the Penrose process and superradiance,  is to have
a massive particle or field scattering off a black
hole, where the mass provides an effective confining
potential. Alternatively, one may consider a massless particle or
field in a black
hole spacetime with small extra dimensions, which generate an
effective mass for the process and thereby act as a confining
mechanism. 

In the Penrose process,
for electrically charged Reissner-Nordstr\"om black holes,
one case that has been studied
is the  disintegrational 
Penrose process in its recursive form,
i.e.,
one has a recursive succession of particle decays in the ergoregion,
with the Penrose process occurring at each decay.
This has been studied for asymptotically flat spacetimes
with a mirror, and for asymptotically AdS spacetimes
with and without mirror. Depending on the conditions 
one gets black hole factories or black hole bombs \cite{paper1}.
In superradiance, for electrically charged Reissner-Nordstr\"om black
holes with a mirror and for Kerr black holes also with a mirror,
the black hole bomb has been studied in several
works cited above.

\subsection{Aim and organization}

It is of interest to understand what is the endpoint of these
recursive systems. For that, one must include backreaction effects,
i.e., one has to consider that the parameters of the black hole,
namely its mass, electric charge, or angular momentum, change along
the recursive process.  For the Penrose process back reaction
considerations in recursive processes have not been done.  For the
superradiance process it was found that, in the electric charged case,
linear instabilities in scalar fields in Reissner-Nordstr\"om-AdS
black holes evolve to a hairy black hole nonlinearly.
Under the AdS/CFT correspondence this phenomenon
can be compared to some form of superconducting transition.

Our aim is to investigate backreaction effects in a recursive Penrose
process occurring in a Reissner-Nordstr\"om-AdS black hole spacetime,
i.e., an electrically charged black hole with a negative cosmological
constant in general relativity, and to determine the endpoint of this
process. A recursive chain of particle decays backreacts on the black
hole, leading to changes in its mass and charge throughout the
evolution. Different outcomes arise
depending on three factors: the initial charge of the black hole,
the initial charge of the decaying particle, and the distribution of
charge among the particles emitted in the decay chain. Possible
endpoints include a charged black hole whose charge equals the sum of
the initial black hole charge and the initial particle charge, a
neutral black hole accompanied by a charged particle at rest at the
location of the first decay, or a black hole with a very small
residual charge, where the
approximation used breaks down and the system evolves in
some other way. We consider the motion of charged
particles in Reissner-Nordstr\"om-AdS black hole spacetimes, following
\cite{olivaresetal}.

This work is organized as follows. In Sec.~\ref{sec:equations}, the
main equations describing charged particle motion in a
Reissner-Nordstr\"om-AdS black hole spacetime are presented, the
equations for the turning points are derived, and particle confinement
in AdS  is presented.
In Sec.~\ref{sec:conditions}, the important assumptions
related to the decays of the particles are stated and the main
equations regarding backreaction on black hole's mass and charge are
derived. Two different cases arise
depending on the value of an index that gives the
number of decays for which
the black hole's electric charge is zero.
One case is when the
index is an integer, the other case is
when the index is not an integer.
This index
depends on the initial electric charges of the black hole and the
decaying particle. 
In
Sec.~\ref{sec:confined}, the two cases are
analyzed in terms of the endpoint of the process, the
energy extraction, and the possibility of a black hole bomb.
In Sec.~\ref{sec:concl}, we conclude,
contemplating the different possibilities that backreaction introduces
and comparing our results to the ones obtained in previously
neglecting backreaction.  In Appendix \ref{app:mirror} we comment on
the possibility of putting a reflective mirror at some radius and show
that having a mirror makes no difference for the results.
We use geometric
units in which the gravitational constant $G$, and the speed of light
$c$ are set to one, $G=1$ and $c=1$.

\newpage

\section{Line element, equations of motion, turning points, and
particle confinement}
\label{sec:equations}

\subsection{Line element}

We consider a Reissner-Nordsr\"om-AdS black hole spacetime, with the
line element  $ds$
for the spacetime interval expressed in spherical
coordinates $(t,r,\theta,\phi)$ as follows
\begin{equation}
 ds^2 = - f\left(r\right) dt^2 +
\frac{dr^2}{f\left(r\right)} + r^2
\left( d\theta^2+\sin^2\theta\, d\phi^2\right),
\quad\quad\quad
f\left(r\right) = \frac{r^2}{l^2} + 1 -
\frac{2M}{r} +\frac{Q^2}{r^2}\,,
 \label{eq:f_RN_AdS}
\end{equation}
where in this context, $f(r)$ denotes the metric function
characterizing the spacetime geometry, with $M$ 
representing the mass
of the black hole, $Q$ its electric charge,
and the parameter $l = \sqrt{- \frac3{\Lambda}}$
defines the length scale associated with the
negative cosmological constant $\Lambda$. 
The coordinate ranges are
$-\infty<t<\infty$, $r_+<r<\infty$,
$0\leq\theta\leq\pi$, and $0\leq\phi<2\pi$. Here, 
$r_+$
denotes the event horizon radius of the black hole, which corresponds
to the largest root of the function $f(r)$, determined by solving
$f(r)=0$, 
so that 
\begin{equation}
r_+=r_+(M,Q,l)\,,
\label{eq:r+def}
\end{equation}
for some definite expression of
$r_+(M,Q,l)$.
The metric function $f(r)$ also vanishes at another radius $r_-$,
corresponding to the inner or Cauchy horizon. This allows the function
$f(r)$ to be factorized in the form
$f(r)= \frac{1}{r^2}(r-r_+)(r-r_-)\left(\frac{r^2}{l^2} +
\frac{r_+ + r_-}{l^2}
r + 1+\frac{r_+^2 + r_-^2 + r_+ r_-}{l^2}\right)$.
The full expressions for $r_+$ and $r_-$ in terms of the black hole
parameters $M$, $Q$, and $l$ are given in \cite{paper1}.
Since our focus is
on the dynamics of particles in the region exterior to the black hole,
i.e., $r>r_+$, the inner horizon will not be relevant to our
analysis.
The electric potential $\varphi$
associated with the black hole is given by
\begin{equation}
\varphi=\frac{Q}{r}\,,
\label{eq:electricpot}
\end{equation}
representing a Coulomb-type potential due to the black hole's 
electric charge $Q$.

\subsection{Equations of motion}

To analyze the dynamics of a charged particle in a
Reissner-Nordstr\"om-AdS spacetime, one can employ a Lagrangian
formalism that incorporates both gravitational and electromagnetic
interactions. For a particle of mass $m$
and electric charge $e$, the
equations of motion follow from the Lagrangian
\begin{equation}
2L = - f\left(r\right) \dot{t}^2 +
\frac{\dot{r}^2}{f\left(r\right)} + r^2 \dot{\phi}^2 -
\frac{2 {\frac{e}{m}} Q}{r} \dot{t}\,,
\label{eq:lagrangian}
\end{equation}
where $f(r)$ is the black hole metric function, and the final
term represents the coupling of the particle to the electric potential
$\varphi=\frac{Q}{r}$.  Here, the dot denotes differentiation with
respect to the particle's proper time $\tau$, which is related to the
spacetime interval via $d\tau^2=-ds^2$.

The equations of motion for a charged particle in the
Reissner-Nordstr\"om-AdS spacetime can be derived directly from the
Lagrangian in Eq.~\eqref{eq:lagrangian} using the
corresponding Euler-Lagrange
equations. Starting with the temporal coordinate $t$, the 
Euler-Lagrange equation reads
 $\frac{d \;}{d
\tau}\left(\frac{\partial L}{\partial\dot t}\right)=0$,
which implies the 
following equation,
$\frac{d
\;}{d \tau}\left(f{\dot t}+\frac{ {\tilde e} Q}{r} \right)=0$.
Integrating, one finds a conserved quantity $E$, interpreted 
as the energy of the particle, namely, 
$f{\dot t}+\frac{ {\frac{e}{m}} Q}{r}
=\frac{E}{m}$.
Then, 
the time component of the four-velocity is given by
${\dot t}=\frac{E-\frac{ e Q}{r}}{mf}$.
Now considering purely radial motion,
i.e., zero angular momentum for the particle, 
the equation for the radial coordinate $r$ yields
$\dot{r}^2 = \frac{\left( E-\frac{ e Q}{r}\right)^2}{m^2}-f$.
It is 
now convenient to define the
time and radial components of the four-momentum as
 $p^t\equiv
m \dot{t}$ and $p^r\equiv m \dot{r}$, respectively.
Then, one can write
\begin{equation}
p^t\equiv m \dot{t} = \frac{X}{f},
\quad\quad\quad\quad
p^r \equiv m \dot{r} = \sigma P\,,
\label{eq:motion_t_penrose}
\end{equation}
with the auxiliary definitions,
\begin{equation}
X =   E - \frac{ e Q}{r}, \quad\quad\quad\quad P =
  \sqrt{X^2 - m^2 f},
    \label{eq:motion_r_penrose}
\end{equation}
and
$\sigma$
in Eq.~\eqref{eq:motion_t_penrose} is
given by 
$\sigma
= \pm1$,
with
$\sigma = + 1$
corresponding to outward radial motion and
$\sigma = - 1$ to inward motion.
Finally, the condition $\dot{t} > 0$, 
which ensures that motion progresses forward in time, 
implies $X > 0$ outside the event horizon.

\subsection{Turning points}

\begin{figure}[h]
    \centering
    \includegraphics[width = 0.42\textwidth, height =
    0.2\textheight]{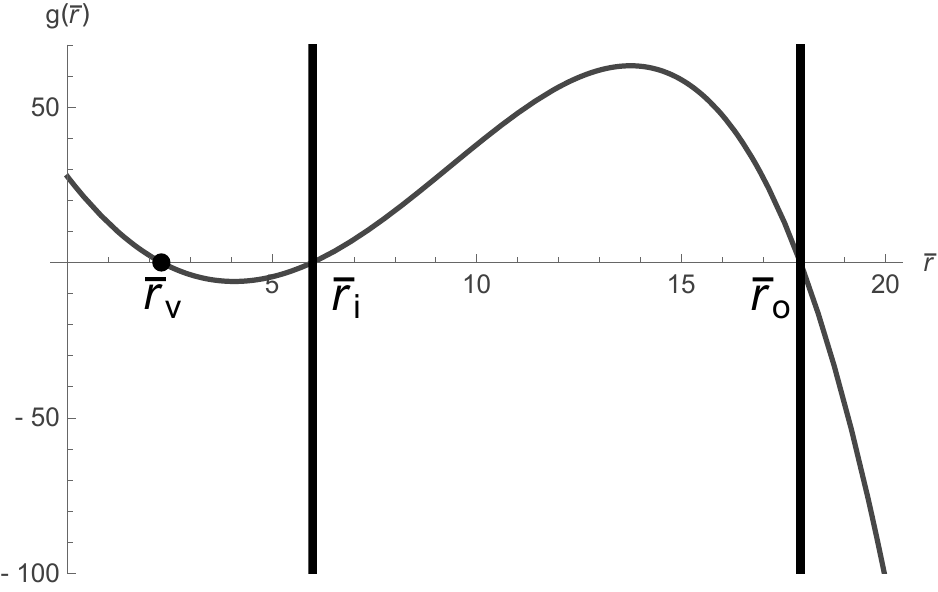}
\caption{
It is plotted the function $g(r) \equiv - \frac{r^4}{M^2 \, l^2} +
\left(\frac{E^2}{m^2} - 1\right) \frac{r^2}{M^2} + 2 \left(1 -
\frac{EeQ}{m^2M}\right) \,\frac{r}{M} + \left(\frac{e^2}{m^2} -
1\right) \frac{Q^2}{M^2}$ defined from
Eq.~\eqref{eq:turning_points_penrose} as a function of the 
radius$r$. The turning points correspond to $\dot
r=0$, i.e., $g(r)=0$.  They are the radius
$ r_{\rm v}$ (black dot) in the vicinity
of the event horizon and
of no interest to us, the internal
turning point with radius $ r_{\rm i}$ (first vertical black
line), and the external turning point with radius $ r_{\rm o}$
(second vertical black line).
The quantities used are unitless,
with the black hole mass $M$ serving as the rescaling quantity.
The rescaled quantities are $\bar
Q=\frac{Q}{M}$, $\bar E =\frac{E}{M}$, $\bar e =\frac{e}{M}$, $\bar m
=\frac{m}{M}$, $\bar l =\frac{l}{M}$,
and $\bar r =\frac{r}{M}$, The values used are $\bar
Q=0.78$, $\bar E = 0.00018$, $\bar e = 0.00068$, $\bar m =0.0001$, and
$\bar l =15.3$.
For these values of the rescaled quantities, the
event horizon is at radius 
$\bar r_+ =1.60283$,
and the 
rescaled turning point radii have values $ \bar r_{\rm v} = 2.29101$,
$\bar r_{\rm i} = 5.9899$, and $\bar r_{\rm o} = 17.9198$.
We are interested in the motion
of particles between $\bar r_{\rm i}$ and $\bar
r_{\rm o}$.
}
\label{fig:roots}
\end{figure}

The turning points $r_{\rm t}$ of an electrically charged
particle with mass
in a Reissner-Nordstr\"om-AdS spacetime correspond to the
radial positions where the particle's radial velocity vanishes. These
points $r_{\rm t}$ mark the extrema of the particle's radial motion
and can be found by solving the equation
\begin{equation}
\dot r=0\,.
    \label{eq:turning_points_generic}
\end{equation}
From Eq.~\eqref{eq:motion_t_penrose} this means $P=0$, and 
making use of Eq.~\eqref{eq:motion_r_penrose}
one gets 
\begin{equation}
- \frac{r^4}{M^2 \, l^2} +
 \left(\frac{E^2}{m^2} - 1\right)
 \frac{r^2}{M^2} + 2 \left(1 - 
 \frac{EeQ}{m^2M}\right) \,\frac{r}{M} +
 \left(\frac{e^2}{m^2} - 1\right) \frac{Q^2}{M^2}=0\,.
    \label{eq:turning_points_penrose}
\end{equation}
This equation has at most four solutions,
the full four solutions arise 
provided that $e$ is sufficiently
large and has the same sign of $Q$. One root has negative
radius and
therefore has no physical meaning. Another root corresponds to a
turning point  in the
vicinity of the event horizon, call its radius $r_{\rm v}$, it is 
a turning point 
for particles coming outward from near the
horizon which then return,
are absorbed by the black hole, and do not escape again,
and so is a turning point of no
interest in this study. The first root of
interest, is an inner turning point for
incoming particles, which we denote as $r_{\rm i}$.
The second root of interest
is an
outer turning point, which we denote as $r_{\rm o}$.
We will focus on
particle motion between the inner and the outer turning points,
$r_{\rm i}$ and $r_{\rm o}$, respectively.  In Fig.~\ref{fig:roots},
the function $g(r) \equiv - \frac{r^4}{M^2 \, l^2} +
 \left(\frac{E^2}{m^2} - 1\right)
 \frac{r^2}{M^2} + 2 \left(1 - 
 \frac{EeQ}{m^2M}\right) \,\frac{r}{M} +
 \left(\frac{e^2}{m^2} - 1\right) \frac{Q^2}{M^2}$,
 which is defined
 from Eq.~\eqref{eq:turning_points_penrose}, is
plotted as a function of the rescaled radius
$\frac{r}{M}$, for a specific choice of
parameters. The three mentioned positive roots of $g(r)=0$ are
highlighted in the figure.

\subsection{Particle confinement in a Reissner-Nordstr\"om-AdS black
hole spacetime}

We study a recursive chain of particle decays occurring at the inner
turning point $r_{\rm i}$. This chain can be sustained in
two different ways, either by
 using the turning point $r_{\rm o}$
or by
placing a reflective mirror at a radius
$r_{\rm m}$.  In both
scenarios, the originally emitted particles are reflected back to the
decay location $r_{\rm i}$.  Neglecting backreaction effects, it was
shown in \cite{paper1} that the first setup
leads to a black hole energy factory, while the second
set up 
can lead to the formation
of a black hole bomb.
In this work, we include backreaction effects to better
understand the underlying dynamics. In particular, we will show
that there is effectively no distinction between the confinement due
to
the turning point $r_{\rm o}$
or 
a reflective mirror, once
backreaction is taken into account, see Appendix
\ref{app:mirror}. Therefore, we consider from the beginning a
situation with no mirror, in which particles are confined just due to
the outer turning point $r_{\rm o}$.

\section{Conditions for the decay, backreaction, energy extraction and
particles' mass and electric charge deployment}
\label{sec:conditions}

\subsection{Conditions for the particle decay}

\subsubsection{Conservation laws for the first particle decay}

We consider the decay of particle 0 at the turning point $r_{\rm i}$ into
particles 1 and 2, subject to the conservation of energy
$E$, momentum $p$,
and electric charge $e$, so that
\begin{equation}
E_0 = E_1 + E_2\,,\quad\quad\quad\quad {\rm at} \;\, r=r_{\rm i},
\label{eq:energy_penrose}
\end{equation}
\begin{equation}
p_0^r = p_1^r + p_2^r\,,\quad\quad\quad\quad {\rm at} \;\, r=r_{\rm i},
\label{eq:momentum_penrose}
\end{equation}
\begin{equation}
e_0 = e_1 + e_2\,,\quad\quad\quad\quad {\rm at} \;\, r=r_{\rm i}.
\label{eq:charge_penrose}
\end{equation}
These conservation laws imply
\begin{equation}
 X_0 = X_1 + X_2\,,\quad\quad\quad\quad {\rm at} \;\, r=r_{\rm i}.
\label{eq:X_penrose}
\end{equation}

We assume that particle 0 decays from rest at the turning point $r_{\rm i}$,
so that
$\dot r_{0}=0$ at  $r=r_{\rm i}$.
From the
definition of momentum, Eq.~\eqref{eq:motion_t_penrose}, and the
conservation of momentum, Eq.~\eqref{eq:momentum_penrose}, we obtain
 at  $r=r_{\rm i}$ that $m_0 \dot r_0= m_1 \dot r_1+
m_2 \dot r_2$,
where $\dot r_1$ and $\dot r_2$
are the radial velocities of particles 1 and 2 at the inner 
turning point $r_{\rm
i}$. 
Since 
$\dot r_0=0$, 
it follows that
 $0=m_1 \dot r_1+ m_2 \dot r_2$.
To simplify, we consider the special case in 
which particles 1 and 2 emerge from the decay at rest, i.e., 
 $\dot r_1=0$ and $\dot r_2=0$, 
which satisfies momentum conservation and simplifies 
subsequent calculations. Thus, we obtain,
\begin{equation}
\dot r_0=0\,,\quad\quad
\dot r_1=0\,,\quad\quad
\dot r_2=0\,,\quad\quad\quad\quad {\rm at} \;\, r=r_{\rm i}.
\label{eq:momentumspeciic_penrose}
\end{equation}
From Eq.~\eqref{eq:motion_t_penrose} this is the same as 
$P_0=0$, $P_1=0$, $P_2=0$ at $r=r_{\rm i}$. Let us see
how to proceed from here
for the subsequent particle decays.


\subsubsection{Conservation laws for subsequent particle decays}

After the decay of particle 0, we assume that particle 1 moves
radially toward the black hole, while particle 2 moves radially
outward. Particle 2 is then reflected back at
the outer turning point $r_{\rm o}$.  Upon returning to the inner
turning point $r_{\rm i}$, particle 2 has zero velocity there. This is
the case because we have assumed that after the decay of particle 0,
particle 2 is at rest, as stated in
Eq.~\eqref{eq:momentumspeciic_penrose}. Therefore, $r_{\rm i}$ must be
a solution of Eq.~\eqref{eq:turning_points_penrose} for the mass,
charge and energy of particle 2. Since those do not change after the
decay, and we assume that the black hole mass and charge change only
after particle 2 reaches $r_{\rm i}$, this position remains a turning
point throughout the process.  Then, once particle 2 reaches $r_{\rm
i}$ it is assumed that particle 2 undergoes a decay analogous to that
of particle 0, i.e., it decays into particle 3, which moves inward,
and particle 4, which is emitted outward.

This process initiates a repeating chain of decays. Each even-numbered
particle $2n$ decays at $r_{\rm i}$ into an odd particle $2n+1$,
which falls into the black hole, and an even particle $2n+2$, which is
emitted outward. The latter is subsequently reflected and returns to
$r_{\rm i}$, where it decays again. This sequence can, in principle,
continue indefinitely for $n=0,1,2,...$.  The expressions derived
earlier for the initial decay can be generalized to describe any step
in this decay chain.  Each decay is assumed to conserve energy,
momentum, and electric charge. Thus, we can write
\begin{equation}
E_{2n}=E_{2n+1}+E_{2n+2}\,,\quad\quad\quad\quad
{\rm at} \;\, r=r_{\rm i},
\label{eq:consEdecay2n}
\end{equation}
\begin{equation}
p_{2n}^r = p_{2n+1}^r + p_{2n+2}^r\,,
\quad\quad\quad\quad {\rm at} \;\, r=r_{\rm i},
\label{eq:momentum_penrose2n}
\end{equation}
\begin{equation}
e_{2n} = e_{2n+1} + e_{2n+2}\,,
\quad\quad\quad\quad {\rm at} \;\, r=r_{\rm i}.
\label{eq:charge_penrose2n}
\end{equation}
These conservation laws directly yield the following relation,
\begin{equation}
X_{2n} = X_{2n+1} + X_{2n+2}\,,
\quad\quad\quad\quad {\rm at} \;\, r=r_{\rm i}.
\label{eq:X_penrose2n}
\end{equation}

When particle $2n$ reaches the turning point $r_{\rm i}$ it has zero
radial velocity $\dot r_{2n}=0$.  This must be the case since we are
considering decays analogous to the first one, i.e., even particles
decay into two particles with initially zero velocity. Thus, provided
that particle $2n$ was at rest immediately after the decay at $r_{\rm
i}$, it must reach this point again with zero radial velocity, since
Eq.~\eqref{eq:turning_points_penrose} is satisfied at $r_{\rm i}$,
both immediately after the decay and when particle $2n$ returns to the
decay point.  There at $r_{\rm i}$ particle $2n$ then decays from
rest.  From the conservation of momentum,
Eq.~\eqref{eq:momentum_penrose2n}, and using the definition of momentum
provided in Eq.~\eqref{eq:motion_t_penrose}, we obtain the relation
$m_{2n} \, \dot r_{2n}= m_{2n+1}\,\dot r_{2n+1}+m_{2n+2}\,\dot
r_{2n+2}$, where $\dot r_{2n+1}$ and $\dot r_{2n+2}$ denote the radial
velocities of particles ${2n+1}$ and ${2n+2}$ at the point $r_{\rm
i}$, respectively.  Since the decay occurs at the inner turning point,
we have $\dot r_{2n}=0$, which leads to $0= m_{2n+1}\,\dot r_{2n+1}+
m_{2n+2}\,\dot r_{2n+2}$.  To further simplify the analysis, we assume
that particles ${2n+1}$ and ${2n+2}$ are produced at rest immediately
after the decay.  Thus,
\begin{equation}
\dot r_{2n}=0\,,\quad\quad
\dot r_{2n+1}=0\,,\quad\quad
\dot r_{2n+2}=0\,,
\quad\quad\quad\quad {\rm at} \;\, r=r_{\rm i}.
\label{eq:momentumspeciic_penrose2n}
\end{equation}
We see that 
this specific scenario satisfies momentum conservation 
and significantly simplifies the calculations.
From Eq.~\eqref{eq:motion_t_penrose} the equations given 
in Eq.~\eqref{eq:momentumspeciic_penrose2n}
imply that 
$P_{2n}=0$, $P_{2n+1}=0$, $P_{2n+2}=0$
at $r=r_{\rm i}$.

\subsection{Backreaction on the black hole's mass $M$, electric
charge $Q$, and metric potential $f$}

In the recursive chain of decays, every odd-numbered particle falls
into the black hole, thereby altering its mass $M$ and electric charge
$Q$. After $n+1$ decays, a total of $n$ particles have fallen into
the black hole. Consequently, the black hole's mass and charge now
depend on $n$, the number of infalling particles.

Just after the $(n+1)$-th decay and before the next odd particle,
labeled $2n+1$, is absorbed, the black hole has mass
\begin{equation}
M_n = M_0 +
E_1 + E_3 + ...+E_{2k+1}+... ,
\label{eq:bh_mass_n1}
\end{equation}
where $M_0$ is the initial mass  of the black hole, 
$E_k$
is the energy of the $k$-th particle, and
$k=0,1,2,...,n-1$.

Just after the $(n+1)$-th decay and before the next odd particle,
labeled $2n+1$, is absorbed, the black hole has electric charge
\begin{equation}
Q_n = Q_n = Q_0 +
e_1 + e_3 + ...+e_{2k+1}+... \,.
\label{eq:bh_charge_n1}
\end{equation}
where $Q_0$ is the initialcharge  of the black hole, 
$e_k$
is the electric charge of the $k$-th particle, and
$k=0,1,2,...,n-1$.

Taking into account these backreaction effects, the metric potential
introduced in Eq.~\eqref{eq:f_RN_AdS} is no longer a 
fixed function of $r$ throughout
the decay chain, as it now depends on the mass $M_n$ and charge
$Q_n$ of the black
hole, both of which evolve during the recursive process. Consequently,
after $n+1$ decays, i.e., just before the odd particle $2n+1$ falls
into the black hole, the metric potential now denoted 
by $f_n$ is given by
\begin{equation}
f_n\left(r\right) = \frac{r^2}{l^2} + 1 -
\frac{2M_n}{r} +\frac{Q_n^2}{r^2}\,,
 \label{eq:f_RN_AdS_n}
\end{equation}
where $M_n$ and $Q_n$
are defined in Eqs.~\eqref{eq:bh_mass_n1}
and \eqref{eq:bh_charge_n1}, respectively.
Thus, we account for the
influence of the absorbed particles on the metric potential by
incorporating their contributions to the black hole's mass and
charge. However, we neglect their effect on any possible deviations
from spherical symmetry.

\subsection{Conditions for energy extraction}

\subsubsection{Energy extraction in the first particle decay}

Since the initial mass and charge of the
black hole are written as $M_0$ and $Q_0$, respectively,
we have that the metric potential given in 
Eq.~\eqref{eq:f_RN_AdS_n} in this initial state is
$f_0\left(r\right) = \frac{r^2}{l^2} + 1
- \frac{2M_0}{r} +\frac{Q_0^2}{r^2}$.  
We have seen that  the conditions of
Eq.~\eqref{eq:momentumspeciic_penrose} imply $P_0=0$, $P_1=0$,
and 
$P_2=0$ at $r_{\rm i}$,
so Eq.~\eqref{eq:motion_r_penrose} gives for the first decay
the result
\begin{equation}
X_0
= m_0 \sqrt{f_0\left(r_{\rm i}\right)}\,,\quad\quad
X_1
= m_1 \sqrt{f_0\left(r_{\rm i}\right)}\,,\quad\quad
X_2
= m_2 \sqrt{f_0\left(r_{\rm i}\right)}\,.
\label{eq:Xi_penrose}
\end{equation}
From the conservation condition
given in Eq.~\eqref{eq:X_penrose}, we have
that mass must be conserved in the first decay, i.e.,
\begin{equation}
m_0 = m_1 + m_2\,.
\label{eq:masscons_penrose}
\end{equation}
Then, from Eqs.~\eqref{eq:motion_r_penrose} and
\eqref{eq:Xi_penrose}, one obtains the following expressions for the
energy of particles 0, 1, and 2,
\begin{equation}
 E_0 = m_0 \sqrt{f_0\left(r_{\rm i}\right)} +
 \frac{e_0Q_0}{r_{\rm i}},\quad\quad
 E_1 = m_1 \sqrt{f_0\left(r_{\rm i}\right)} +
 \frac{e_1Q_0}{r_{\rm i}},\quad\quad
 E_2 = m_2 \sqrt{f_0\left(r_{\rm i}\right)} +
 \frac{e_2Q_0}{r_{\rm i}}.
    \label{eq:E_1_penrose}
\end{equation}
Since, 
from Eq.~\eqref{eq:charge_penrose}, electric charge is conserved, i.e.,
$e_0 = e_1 + e_2$,
clearly, the energies given in Eq.~\eqref{eq:E_1_penrose}
satisfy the energy conservation in the first decay,
$E_0=E_1+E_2$, as required by
Eq.~\eqref{eq:energy_penrose}.

To enable energy extraction via the Penrose process in this decay, the
energy of particle 1 must be negative, $E_1 < 0$. 
This ensures that the
energy of the emitted particle, particle 2, 
is greater than that of the
initial particle 0, which has decayed
at radius  $r_{\rm i}$. Consequently, the
decay point $r_{\rm i}$ must lie within the electric ergoregion, where
 by
definition
particles with a given electric charge can possess negative energy.
By appropriately choosing the electric charge $e_1$
 of particle 1, one
can guarantee that $r_{\rm i}$ remains inside the ergoregion. From
Eq.~\eqref{eq:E_1_penrose}, it follows that particle 1 has negative
energy, i.e., $r_{\rm i}$
lies inside the ergoregion, if and only if the
following two conditions are satisfied:
1. $e_1 = - \lvert
e_1 \rvert$, 2. $\lvert e_1 \rvert > \frac{r_{\rm
i}}{Q_0} \sqrt{f_0\left(r_{\rm i}\right)}\, m_1$.
Here, we assume without loss of generality that $Q_0 > 0$. This
assumption will be maintained throughout. If instead the black hole
initially had negative charge $Q_0 < 0$, energy extraction would still
be possible, provided that the sign of $e_1$
 is reversed
accordingly. In general, energy extraction requires the condition
 $Q_0 e_1 < 0$.
In summary, the following conditions must be met to make certain
that there is energy
extraction during the first decay and that $r_{\rm i}$
lies within the
ergoregion,
\begin{equation}
E_{1} <0\,,
\quad\quad\quad
E_{2}> E_{0}\,,
\quad\quad\quad\quad {\rm at} \;\, r=r_{\rm i}.
    \label{eq:conditiondecay1initial}
\end{equation}
which implies 
\begin{equation}
e_{1} = - \lvert e_{1} \rvert\,,
\quad\quad\quad
\lvert e_{1} \rvert > \gamma_0 m_1\,,\quad\quad
\gamma_0 \equiv\frac{r_{\rm i}}{Q_0} \sqrt{f_0\left(r_{\rm i}\right)}\,.
\label{eq:conditiondecay2initial}
\end{equation}
The latter inequality shows that it is sufficient to choose a suitable
electric charge for particle 1 to impose energy extraction in the
first decay.

\subsubsection{Energy extraction in subsequent particle decays}

Just after the $(n+1)$-th decay of the particle $2n$, 
and before the next odd particle
 $2n+1$ is absorbed and the 
next even particle $2n+2$ is emitted, the 
black hole has mass
$M_n$ and electric charge $Q_n$.
The metric potential becomes
$f_n\left(r\right) = \frac{r^2}{l^2} + 1
- \frac{2M_n}{r} +\frac{Q_n^2}{r^2}$, see Eq.~\eqref{eq:f_RN_AdS_n}.  
We have seen that the conditions of
Eq.~\eqref{eq:momentumspeciic_penrose2n}
imply that at $r_{\rm i}$
one has
$P_{2n}=0$, $P_{2n+1}=0$, $P_{2n+2}=0$ at $r_{\rm i}$.
So, just after the $(n+1)$-th decay of the particle $2n$,
the definition in 
Eq.~\eqref{eq:motion_r_penrose} yields
\begin{equation}
X_{2n}
= m_{2n} \sqrt{f_n\left(r_{\rm i}\right)}\,,\quad\quad
X_{2n+1}
= m_{2n+1} \sqrt{f_n\left(r_{\rm i}\right)}\,,\quad\quad
X_{2n+2}
= m_{2n+2} \sqrt{f_n\left(r_{\rm i}\right)}\,.
\label{eq:Xi_penrose2n}
\end{equation}
Then, the conservation condition $X_{2n} = X_{2n+1} + X_{2n+2}$,
see Eq.~\eqref{eq:X_penrose2n},
yields directly 
that mass is
conserved in each decay, i.e.,
\begin{equation}
m_{2n} = m_{2n+1} + m_{2n+2}\,.
\label{eq:masscons_penrose2n}
\end{equation}
It follows from  Eqs.~\eqref{eq:motion_r_penrose}
and ~\eqref{eq:Xi_penrose2n}, 
that the energies of the particles just before 
and just after the  $(n+1)$-th decay
are 
\begin{equation}
 E_{2n} = m_{2n} \sqrt{f_n\left(r_{\rm i}\right)} +
 \frac{e_{2n}Q_n}{r_{\rm i}},\quad\;
 E_{2n+1} = m_{2n+1} \sqrt{f_n\left(r_{\rm i}\right)} +
 \frac{e_{2n+1}Q_n}{r_{\rm i}},
\quad\;
E_{2n+2} = m_{2n+2} \sqrt{f_n\left(r_{\rm i}\right)} +
 \frac{e_{2n+2}Q_n}{r_{\rm i}}.
    \label{eq:E_odd_penrose}
\end{equation}
These equations
imply energy conservation in each decay, i.e.,
$E_{2n}=E_{2n+1}+E_{2n+2}$, as imposed in 
Eq.~\eqref{eq:consEdecay2n}.

To enable energy extraction via the Penrose process during the decay
of particle $2n$, the energy of the resulting particle $2n+1$
must
be negative, $E_{2n+1}<0$. This condition ensures that the energy of
the emitted particle $2n+2$
exceeds that of the decaying particle
$2n$.
Equivalently, for energy extraction to occur in the $(n+1)$th
decay, which takes place at the turning point $r_{\rm i}$, 
this point must
lie within the electric ergoregion. This ergoregion is defined as the
region where particle $2n+1$ has negative energy, given its
electric charge $e_{2n+1}$.
Thus, by appropriately choosing the electric charge of particle 
 $2n+1$, the 
turning point $r_{\rm i}$
can be positioned inside the ergoregion.
From Eq.~\eqref{eq:E_odd_penrose}, 
energy extraction is possible if and only if two conditions are met:
1. $e_{2n+1} = -
\lvert e_{2n+1} \rvert$, i.e., the electric charge is negative, 
2.  $\lvert e_{2n+1} \rvert >
\frac{r_{\rm i}}{Q_n} \sqrt{f_n\left(r_{\rm i}\right)}\, m_{2n+1}$, 
i.e.,  the magnitude of the charge has
to be sufficiently large.
Here, it is assumed that $Q_n>0$, as discussed earlier.
In summary, energy extraction in the decay of particle $2n$
requires
that the turning point $r_{\rm i}$
lie within the electric ergoregion, which
can be guaranteed by appropriately choosing the charge $e_{2n+1}$ to
satisfy the conditions above.
Thus, 
\begin{equation}
E_{2n+1} <0\,,
\quad\quad\quad
E_{2n+2}> E_{2n}\,,
\quad\quad\quad\quad {\rm at} \;\, r=r_{\rm i},
\label{eq:conditiondecay1}
\end{equation}
which implies 
\begin{equation}
e_{2n+1} = - \lvert e_{2n+1} \rvert\,,
\quad\quad\quad
\lvert e_{2n+1} \rvert > \gamma_n m_{2n+1}\,,
\quad\quad
\gamma_n \equiv
\frac{r_{\rm i}}{Q_n} \sqrt{f_n\left(r_{\rm i}\right)}\,.
\label{eq:conditiondecay2}
\end{equation}
The latter inequality
shows that it is sufficient to choose a suitable electric
charge for particle $2n+1$ to impose energy extraction in the decay of
particle $2n$.
Note that
Eqs.~\eqref{eq:conditiondecay1}
and \eqref{eq:conditiondecay2}
generalize the conditions of 
Eqs.~\eqref{eq:conditiondecay1initial}
and \eqref{eq:conditiondecay2initial}
found for the initial decay.

\subsection{Mass and charge conservation
after $n+1$ decays}
\label{sec:twoscenarios}

\subsubsection{Mass and electric charge
of the decaying particles}

After $n+1$ decays the masses
of decaying particles, $m_{2n+1}$ and $m_{2n+2}$, are written as
$m_{2n+1}=\alpha_1 m_{2n}$ and $m_{2n+2} =\alpha_2 m_{2n}$, 
and we assume that 
$\alpha_1$ and $\alpha_2$ are constant parameters that do not
vary along the decay chain. 
Since the masses of the particles must be positive, one has
$\alpha_1>0$ and
$\alpha_2>0$.
Imposing mass conservation in each decay,
Eq.~\eqref{eq:masscons_penrose}, one has
$\alpha_1 + \alpha_2 = 1$, with
$\quad0<\alpha_1<1$ and 
$0<\alpha_2<1$. In brief, we can write
\begin{equation}
m_{2n+1}=\alpha_1 m_{2n},\quad\quad
m_{2n+2} =\alpha_2 m_{2n},\quad\quad
\alpha_1 + \alpha_2 = 1,\quad\quad
0<\alpha_1<1,\quad 
0<\alpha_2<1.
    \label{eq:mass_odd_case1_penroseall}
\end{equation}
Clearly,  this set of equations  displayed in
Eq.~\eqref{eq:mass_odd_case1_penroseall}
provide iterative expressions for
the masses of the decaying particles. These conditions can be solved
to obtain general expressions for the masses of odd and even particles
after $n+1$ decays, namely,
\begin{equation}
m_{2n+1} =\alpha_1 (1-\alpha_1)^n m_{0}\,,
\label{eq:mass2_case1_penrose}
\end{equation}
\begin{equation}
m_{2n+2}=\alpha_2^{n+1} \,m_{0}\,.\quad\quad\quad
\label{eq:mass1_case1_penrose}
\end{equation}

After $n+1$ decays the charges
of decaying particles,
$e_{2n+1}$ and $e_{2n+2}$, are written as
$e_{2n+1}=\beta_1 e_{2n}$ and $e_{2n+2} =\beta_2 e_{2n}$, 
and we assume that 
$\beta_1$ and $\beta_2$ are constant parameters that do not
vary along the decay chain. 
To fulfill the assumptions stated before, it is necessary 
that $\beta_1 < 0$. From 
 electric charge conservation in all the decays,
Eq.~\eqref{eq:charge_penrose}, one has 
$\beta_1 + \beta_2 = 1$, and so $\beta_2 >0$. 
Therefore the parameters $\beta_1$ and $\beta_2$ obey 
$\beta_1 + \beta_2 = 1$
with $-\infty<\beta_1<0$
and $0<\beta_2<\infty$.
In brief, we can write
\begin{equation}
e_{2n+1}=\beta_1 e_{2n},\quad\quad
e_{2n+2} =\beta_2 e_{2n},\quad\quad
\beta_1 + \beta_2 = 1,\quad\quad
-\infty<\beta_1<0,\quad 
0<\beta_2<\infty.
    \label{eq:charge_odd_case2_penroseall}
\end{equation}
Solving these equations iteratively yields
\begin{equation}
 e_{2n+1}= \beta_1 (1-\beta_1)^n\, e_{0}\,,
    \label{eq:charge2_case2_penrose}
\end{equation}
\begin{equation}
 e_{2n+2}= \beta_2^{n+1} e_{0}\,.
    \label{eq:charge1_case2_penrose}
\end{equation}

This decaying process corresponds to the second scenario
considered in \cite{paper1}, for which it was shown that,
under certain conditions and without considering backreaction effects,
a black hole bomb is possible. 
Note also that
it is not necessary for odd particles to have negative energy in the
initial decays, which means that, in such cases, no energy is
extracted during the first few decay events.
But, since the energy of odd particles decreases
along the decay chain, it becomes negative soon, thus
allowing energy
extraction.

\subsubsection{Mass $M_n$ and electric charge $Q_n$ of the black hole
after $n$ decays and the number $n_{\rm c}$.
Two cases: Case 1, $n_{\rm c}$ integer; Case 2, $n_{\rm c}$ not 
integer}
\label{sec:bh_charge}

As odd particles fall into the black hole, its mass $M$ 
changes according
to Eq.~\eqref{eq:bh_mass_n1} which can be written as
\begin{equation}
M_n = M_0 + \sum_{k=0}^{n-1} E_{2k+1}\,.
\label{eq:bh_mass2}
\end{equation}

The electric charge $Q$ of the
black hole also changes due to the accretion of the odd particles
as $Q_{n+1} = Q_n + e_{2n+1}$. After
$n+1$ decays, according to Eq.~\eqref{eq:bh_charge_n1} and
using the expression for electric charge of odd particles obtained in
Eq.~\eqref{eq:charge2_case2_penrose}, the charge of the black hole is
$Q_n = Q_0 + \beta_1 e_0 \sum_{k=0}^{n-1}
\left(1-\beta_1\right)^k$. Summing the geometric series over 
$k$, one gets
\begin{equation}
Q_n  = Q_0 + \sum_{k=0}^{n-1} e_{2k+1}
=Q_n = Q_0 - e_0 \left(\beta_2^n-1\right)\,,
\label{eq:charge_bh_general}
\end{equation}
or
$Q_0 - e_0 \left((1-\beta_1)^n-1\right)$. 
Since $1-\beta_1=\beta_2 > 1$, the charge of the black hole decreases
along the decay chain.

Define now 
the index $n_{\rm c}$ as the value for which the black hole's charge is zero,
$Q_{n_{\rm c}}=0$.
Thus, the index $n_{\rm c}$ is given by the
solution of the equation  $Q_{n_{\rm c}}=0$, i.e.,
\begin{equation}
n_{\rm c} = \frac{\ln{\left(1+\frac{Q_0}{e_0}\right)}}{\ln{\beta_2}}\,.
\label{eq:n_critical}
\end{equation}
It is an important number. 
We can then divide the problem into two cases.
\noindent
Case 1: If $n_{\rm c}$ is an integer number,
the black hole charge
reaches exactly zero after $n_{\rm c} +1$
decays, i.e., $Q_{n_{\rm c}}=0$. In this
situation, the recursive Penrose process terminates naturally.
\noindent
Case 2: If $n_{\rm c}$ is not an
integer,
 the black hole charge
approaches but does not reach zero. The last admissible decay occurs
at $n_{\rm c}^-$, where $n_{\rm c}^-$
is the greatest integer less than $n_{\rm c}$. At
this stage the remaining charge $Q_{n_{\rm c}^-}$
is small and positive. Beyond
this point, the approximations underlying the analysis cease to be
valid, and additional physical mechanisms intervene to halt the
process.

In brief, one has two possible cases. In case 1, $n_{\rm c}$ is integer and
the charge of the black hole vanishes after $n_{\rm c} +1$ decays
and the process stops naturally.  In case
2, $n_{\rm c}$ is noninteger and the charge of the black hole
decreases up to some small positive value.
At this stage the approximation employed here is no
longer reliable. Indeed, a naive continuation of the calculation would
lead to a gross violation of cosmic censorship.
In what follows, we analyze energy extraction in both cases, assuming
particle confinement provided by the outermost turning point $r_{\rm o}$
of
the AdS geometry.

\subsubsection{Generic expressions for the energy of the particles
after $n+1$ decays}
\label{sec:energy_generic_expressions}

Using the recursive relations for the energy of odd and even
particles, see Eq.~\eqref{eq:E_odd_penrose}, the expressions for the
mass and charge of odd and even particles,
Eqs.~\eqref{eq:mass2_case1_penrose}, \eqref{eq:mass1_case1_penrose},
\eqref{eq:charge2_case2_penrose}, and
\eqref{eq:charge1_case2_penrose}, and the equation for the charge of
the black hole after $n+1$ decays, Eq.~\eqref{eq:charge_bh_general},
the following generic expressions for the energy of odd and even
particles after $n+1$ decays can be obtained  for $n < n_{\rm c}$,
\begin{align}
 &E_{2n+1} = \sqrt{f_n\left(r_{\rm i}\right)}\, \alpha_1
 \left(1-\alpha_1\right)^n m_0 - \frac{e_0 \left(Q_0 +
 e_0\right)}{r_{\rm i}} \,\lvert\beta_1\rvert \left(1-\beta_1\right)^n
 + \frac{e_0^2}{r_{\rm i}} \, \lvert\beta_1\rvert
 \left(1-\beta_1\right)^{2n}\,,\quad\quad\quad n < n_{\rm c},
\label{eq:energy1_case2_penrose}
\\
 &
E_{2n+2} = \sqrt{f_n\left(r_{\rm i}\right)} \, \alpha_2^{n+1} \, m_0
 + \frac{e_0 \left(Q_0 + e_0\right)}{r_{\rm i}}\beta_2^{n+1} -
 \frac{e_0^2}{r_{\rm i}} \, \beta_2^{2n+1} \,,
\quad\quad\quad\quad\quad\quad\quad\quad\quad\quad\quad\quad\quad
\quad  n < n_{\rm c},
\label{eq:energy2_case2_penrose}
\end{align}
where  $f_n(r)$ is given in 
Eq.~\eqref{eq:f_RN_AdS_n}.
From it 
we can write $f_n(r_{\rm i})$ 
as
\begin{equation}
f_n\left(r_{\rm i}\right) = \frac{r_{\rm i}^2}{l^2} + 1 -
\frac{2\left(M_0 + \sum_{k=0}^{n-1} E_{2k+1}\right)}{r_{\rm i}}
+\frac{\left(Q_0 - e_0
\left(\beta_2^n-1\right)\right)^2}{r_{\rm i}^2},
 \label{eq:fnnew}
 \end{equation}
where Eqs.~\eqref{eq:bh_mass2} and
\eqref{eq:charge_bh_general} have been used.
Therefore,
$E_{2n+1}$ and  
$E_{2n+2}$
of Eqs.~\eqref{eq:energy1_case2_penrose} and \eqref{eq:energy2_case2_penrose}
depend on $E_{2n-1}$, i.e., the
energy of the previous particle falling into the black hole, and the
corresponding equations must
be solved iteratively. 

Contrarily to the energies obtained without considering backreaction
effects in \cite{paper1}, 
Eqs.~\eqref{eq:energy1_case2_penrose} and \eqref{eq:energy2_case2_penrose}
each contain now $Q_0 + e_0$ instead
of $Q_0$ and include a term of degree $2n$ and $2n+1$, respectively,
which dominates after a sufficiently large number of decays. For the
energy of odd particles, the highest order term introduces a positive
contribution, making $E_{2n+1}$ non-negative after a sufficiently
large number of decays, a phenomenon that does not occur without
considering backreaction effects.  On the other hand, the highest
order term introduces a negative contribution for the energy of even
particles which together with $n_{\rm c}$ does not allow $E_{2n+2}$ to
diverge, as occurred without considering backreaction effects.
Comparing the two last terms of Eq.~\eqref{eq:energy1_case2_penrose}
we see that if $e_0 \beta_2^n\ll Q_0$, the results of \cite{paper1}
are recovered.

After these settings we can now study the two cases mentioned.
The number $n_{\rm c}$ is
 an integer and  the charge of the black hole reaches
exactly a zero value after $n_{\rm c} +1$ decays, and 
the number $n_{\rm c}$ is
not an integer 
and the charge of the black hole becomes small for the closest
integer smaller than $n_{\rm c}$.


\section{Recursive Penrose processes in Reissner-Nordstr\"om-AdS black hole
spacetimes}
\label{sec:confined}


\subsection{Case 1: Integer value of the index $n_{\rm c}$}

\subsubsection{Preamble}

Here we treat the case of an integer value of the index $n_{\rm c}$.  The
index $n_{\rm c}$ defined in Eq.~\eqref{eq:n_critical} gives the value at
which the charge $Q$ of the black hole has a zero value, namely,
$n_{\rm c} = \frac{\ln{\left(1+\frac{Q_0}{e_0}\right)}}{\ln{\beta_2}}$.
For an integer value of $n_{\rm c}$, the charge of the black hole,
decreasing with the number of decays $n$, reaches exactly zero value
for $n=n_{\rm c}$.

\subsubsection{Electric charge of
particles and electric charge of the black hole}

\begin{figure}[h]
\centering
    \includegraphics[width = 0.39\textwidth, height =
    0.2\textheight]{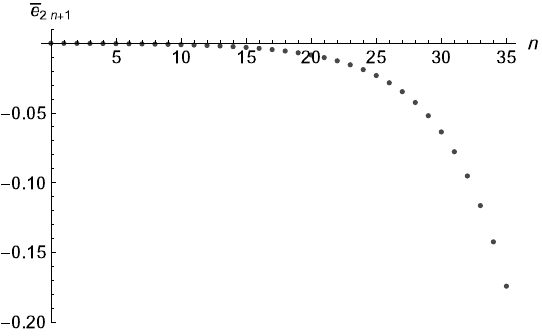}
    \hskip 2cm
    \includegraphics[width = 0.4\textwidth, height =
    0.2\textheight]{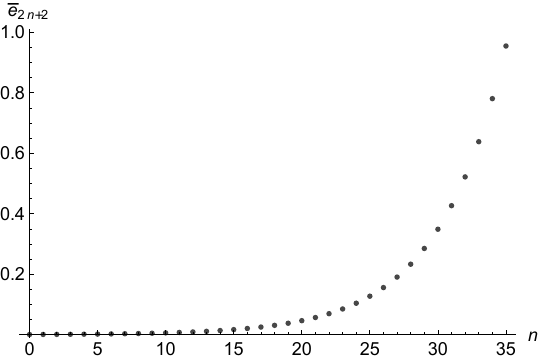}
\caption{
Left: It is plotted the electric charge
$e_{2n+1}$
of the odd particles
as a function of the
number $n$ of the decay. Right: 
It is plotted the electric charge of the even particles
as a function of the
number $n$ of the decay.
Both plots are 
for a recursive Penrose process occurring in a
Reissner-Nordstr\"om-AdS black hole spacetime
when the index
$n_{\rm c}$ is an integer, where $n_{\rm c}$ is the number for which the
charge of the black hole reaches a zero value.
The plots are
for $0 \leq n \leq n_{\rm c}$.
The quantities used
are unitless, with the initial black hole mass $M_0$ serving as the
rescaling quantity.  So the rescaled electric charges $\bar
e_{2n+1}=\frac{e_{2n+1}}{M_0}$ and $\bar
e_{2n+2}=\frac{e_{2n+2}}{M_0}$ are the quantities plotted as a
function of $n$. The other rescaled quantities are $\bar
Q_0=\frac{Q_0}{M}$ and $\bar e_0=\frac{e_0}{M_0}$.  The values used
are $\bar{Q_0}=0.78$, $\bar{e_0}=0.00068$, and $\beta_2 = 1.223$.  For
these values one has that the value of $n$ when the black hole
electric charge is zero is $n_{\rm c} = 35$, for which the
process naturally comes to a halt.}
\label{fig:chargeofparticles}
\end{figure}

To study the evolution of the particles charge we use iteratively
Eqs.~\eqref{eq:charge2_case2_penrose} and
\eqref{eq:charge1_case2_penrose}.  We see that odd particles have a
negative electric charge, whose absolute value increases geometrically
with the number $n$, while even particles have a positive electric
charge, increasing also geometrically. The electric charge of the
particles and the electric charge of the black hole are coupled
through Eq.~\eqref{eq:charge_bh_general}, therefore one must compute
the different charges consistently.  In
Fig.~\ref{fig:chargeofparticles}, it is shown the evolution of
electric charge of the particles $e_{2n+1}$ and $e_{2n+2}$ as a
function of the number of decays $n$.

\begin{figure}[h]
\centering
\includegraphics[width = 0.4\textwidth, height =
0.2\textheight]{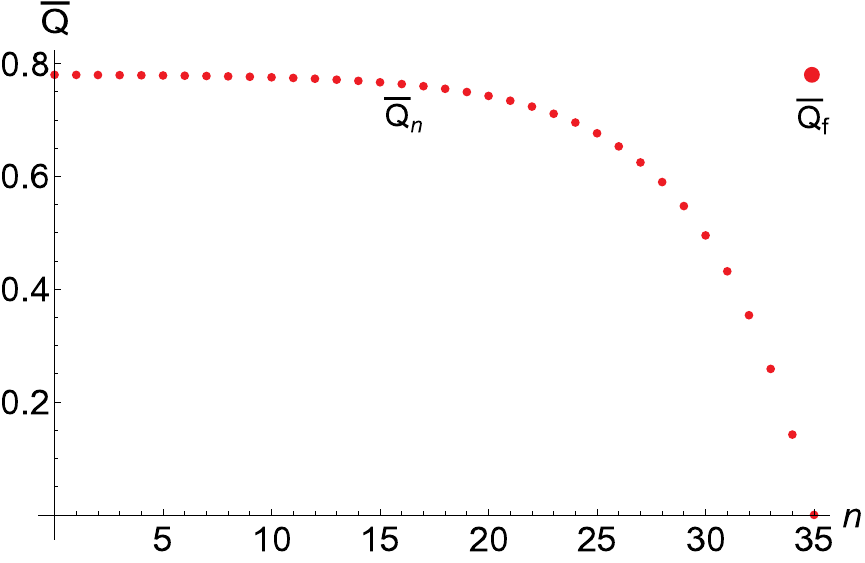}
\caption{
It is plotted the electric charge of the black hole $Q_n$, as a
function of the number $n$ of the decay, for a recursive Penrose
process occurring in a Reissner-Nordstr\"om-AdS black hole
spacetime when the index
$n_{\rm c}$ is an integer, where $n_{\rm c}$ is the number for which the
charge of the black hole reaches a zero value.
The plot is for $0 \leq n \leq n_{\rm c}$.
The final black hole charge $Q_{\rm f}$ is also displayed.
The quantities used are unitless, with the initial black hole mass
$M_0$ serving as the rescaling quantity.  So the rescaled black hole
charge $\bar Q_n=\frac{Q_n}{M_0}$ is the quantity plotted as a
function of $n$. The rescaled final black hole charge $\bar Q_{\rm f}
=\frac{Q_{\rm f}}{M_0}$ is also displayed.  The other
rescaled quantity is $\bar e_0=\frac{e_0}{M_0}$.  The values used are
$\bar{Q_0}=0.78$, $\bar{e_0}=0.00068$, and $\beta_2 = 1.223$.  For
these values one has that the value of $n$ when the black hole
electric charge is zero is $n_{\rm c} = 35$, for which the
process naturally comes to a halt.}
\label{fig:charge}
\end{figure}

To study the evolution of the black hole electric charge we use
Eq.~\eqref{eq:charge_bh_general} iteratively.  After $n_{\rm c}$ decays the
charge of the black hole is zero, $Q_{n_{\rm c}}=0$.  However, the final
charge of the black hole will not be zero.  This is because the
particle $2n_{\rm c}$ will be reflected back from the outer turning point
$r_{\rm o}$ and will fall down directly since the black hole is
momentarily uncharged. Therefore, the final black hole charge is the
electric charge of the particle $2n_{\rm c}$, which is $Q_0+e_0$ by charge
conservation.  So the final black hole electric charge $Q_{\rm f}$ is
\begin{equation}
Q_{\rm f} = Q_0+e_0\,.
    \label{eq:charge_final}
\end{equation}
In Fig.~\ref{fig:charge}, it is shown the charge of the black hole
$Q_n$ decreasing down to zero as a function of $n$ up to $n_{\rm c}$
following Eq.~\eqref{eq:charge_bh_general} and then it gets the final
charge of Eq.~\eqref{eq:charge_final} when particle $2n_{\rm c}$ falls in.

\newpage

\subsubsection{Analysis of the turning points when $n\to n_{\rm c}$ and a
curious configuration}
\label{sec:case2turning}

\centerline{\it\small (a) The turning points}
\vskip 0.2cm

As $n$ approaches the integer $n_{\rm c}$, and so the charge of the black hole
approaches zero, 
i.e., $\frac{Q}{M}\to0$,
Eq.~\eqref{eq:turning_points_penrose} for the turning points
simplifies into
\begin{equation}
- 
\left(
\frac{M^2}{l^2}
\right)
\frac{r^3}{M^3} +
 \left(\frac{E^2}{m^2} - 1\right)
 \frac{r}{M} + 2=0\,,\quad\quad\quad\quad n\to n_{\rm c}
 \,,
    \label{eq:turning_points_penrose_zeroQ}
\end{equation}
which is a third degree polynomial equation
for $r$. Let us discuss the solutions of interest of
Eq.~\eqref{eq:turning_points_penrose_zeroQ}, namely,
the inner and outer turning points, 
$r_{\rm i}$ and $r_{\rm o}$, respectively.

When $n\to n_{\rm c}$,
the turning point $r_{\rm i}$ has an important feature, namely,
it disappears.  Indeed, from
Eq.~\eqref{eq:turning_points_penrose_zeroQ} one finds that after
$n_{\rm c}+1$ decays the turning point $r_{\rm i}$
corresponds to a negative root, and so is of no use.
Therefore, the recursive process must
stop at $n=n_{\rm c}$, with the particles $2n_{\rm c} +1$
and $2n_{\rm c}+2$ falling
both into the black hole. This happens because the black hole is no
longer charged, not being able to repel the positively charged even
particle in order to stop it at the position $r_{\rm i}$.  After
particles $2n_{\rm c}+1$ and $2n_{\rm c}+2$ fall into the black
hole, the black
hole becomes again charged with an electric charge equal to the sum of
the electric charge of these particles, which is given by $Q_0 + e_0$,
as can be found from the conservation of electric
charge.

When $n\to n_{\rm c}$,
the turning point $r_{\rm o}$ still exists and
still has the important feature of being the point where 
particles are reflected back. This outer turning
point $r_{\rm o}$ corresponds to the largest real
solution of Eq.~\eqref{eq:turning_points_penrose_zeroQ}, with its
position depending on  $n\to n_{\rm c}$.
The explicit expression for the position
of the outer turning point can be found in Appendix D of
\cite{paper1}.
Nevertheless is interesting to examine this turning
point a little further
when the
energy per unit mass of the even particle is significantly large,
i.e., $\frac{E}{m}\gg1$.
In this limit of $\frac{E}{m}$
large one can consider analytically two regimes, namely,
$\frac{E}{m} \ll \frac{M}{l}$ and
$\frac{E}{m} \gg \frac{M}{l}$. The intermediate regime of
energies is not analytically tractable but can
then be interpolated between the two limiting regimes.
In
the regime of
$\frac{E}{m} \ll \frac{M}{l}$, the coefficient of the first term
dominates with respect to the coefficient of the second one. Thus,
$r_{\rm o}$ must be smal, such that
$-\left(\frac{M^2}{l^2}\right)\frac{r_{\rm o}^3}{M^3} + 2=0$. This
leads to an outer turning point given by $\frac{r_{\rm
o}}{M}=\sqrt[3]2\left(\frac{l}{M}\right)^\frac{2}{3}$.
But from the definition of $f$ in Eq.~\eqref{eq:f_RN_AdS}, i.e., 
$f\left(r\right) = \frac{r^2}{l^2} + 1 -
\frac{2M}{r} +\frac{Q^2}{r^2}$,
we see that the solution
for the horizon radius $r_+=r_+(M,Q,l)$
of Eq.~\eqref{eq:r+def} is 
$\frac{r_+}{M}=\sqrt[3]2\left(\frac{l}{M}\right)^\frac{2}{3}$.
Thus $r_{\rm o}$ and $r_+$ coincide in the limit of this regime
and the particles are turned back to the hole at the horizon itself.
It is a case of no interest and we do not discuss it further.
In the regime of
$\frac{E}{m} \gg \frac{M}{l}$,
the turning point is further out from $r_+$, this is the case of our
interest.
When
$\frac{E}{m}$ is large and moreover $\frac{E}{m} \gg \frac{M}{l}$, it
can be seen from Eq.~\eqref{eq:turning_points_penrose_zeroQ} that, in
this regime, the position of the
outer turning point must increase, so that the negative term
proportional to $r_{\rm o}^3$ overcomes the positive term proportional
to $\frac{E^2}{m^2} r_{\rm o}$. Indeed, in this regime, the first two
terms of Eq.~\eqref{eq:turning_points_penrose_zeroQ} are much larger
than the third one, so the constant term can be neglected, leading to
$-\left(\frac{M^2}{l^2}\right) \frac{r_{\rm o}^3}{M^3} +
\left(\frac{E^2}{m^2}\right) \frac{r_{\rm o}}{M}=0$. Therefore, in
this regime, one gets $\frac{r_{\rm o}}{M} = \frac{E}{m} \frac{l}{M}$,
which indeed grows with the energy per unit mass $\frac{E}{m}$
of the even particles.
Furthermore,
we see that from
Eq.~\eqref{eq:energy2_case2_penrose}, the energy of even particles
remains finite for all values of $n<n_{\rm c}$.  Moreover, we have found
through numerical calculation
that, for a suitable choice of parameters $\alpha_1$,
$\alpha_2$, $\beta_1$, and $\beta_2$, and initial masses and charges
of the decaying particle and the black hole, specifically, 
 these
quantities have reasonable values and
are not allowed to have enormously large values, 
the position of the outer
turning point is finite.
Note that this finiteness for
the position of the outer
turning point does not happen if one neglects backreaction
effects, as worked out  in \cite{paper1}, since in
such a scenario the energy per unit mass of even particles diverges when $n
\to \infty$, implying that the turning point position increases to
infinity. 
Since backreaction imposes a finite position for the turning point,
particle confinement to a finite volume is ensured.
We have that after the decay of particle
$2n$, with $n<n_{\rm c}$, particle $2n+2$ will bounce back from the
turning point $r_{\rm o}$, reaching $r_{\rm i}$ with zero velocity.
This regime $\frac{E}{m}\gg1$ and $\frac{E}{m} \gg \frac{M}{l}$
is the
case of interest in this work that we analyze in detail
in what follows.

\vskip 0.5cm
\centerline{\it\small (b) A curious configuration}
\vskip 0.2cm

There is a curious possibility. 
Since the outer turning point changes its position
in the process as it depends on $n$,
we can imagine a very fine-tuned  configuration in which the position of
the outer turning point after $n_{\rm c}+1$ decays exactly coincides with
the decay position, i.e., $r_{\rm o}\left(n_{\rm c}\right) = r_{\rm i}$,
where we should bear in mind that $r_{\rm i}$ has now
disappear since the black hole has no electric charge anymore.
In
this case, the particle $2n_{\rm c}$ cannot move outwards immediately after the
decay and remains at rest at the decay position which to avoid
confusion we call $r_{\rm
r}$, the old $r_{\rm i}$. Therefore, in this fine-tuned
situation we end up with an
uncharged black hole, i.e.,
$Q_{\rm f}=Q_{n_{\rm c}}=0$,  with mass $M_{n_{\rm c}}$, and a particle with charge
$Q_0 + e_0$ at rest at the position
$r_{\rm r}$. In brief,
\begin{equation}
Q_{\rm f}=Q_{n_{\rm c}}=0,\quad\quad\quad
e_{2n_{\rm c}}=
 Q_0+e_0\,,\quad\quad\quad r=r_{\rm r},
    \label{eq:curiouss}
\end{equation}
Since the particle is at
rest in a position outside the black hole event horizon, the configuration
can be considered a black hole with small hair. 
We note, however, that the  configuration is unstable,
any perturbation will make the charged particle move inwards
or outwards.

\subsubsection{Energy extraction}

\begin{figure}[h]
    \centering
    \includegraphics[width = 0.42\textwidth, height =
    0.2\textheight]{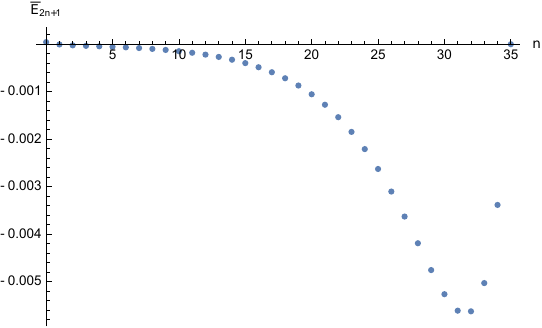}
    \includegraphics[width = 0.42\textwidth, height =
    0.2\textheight]{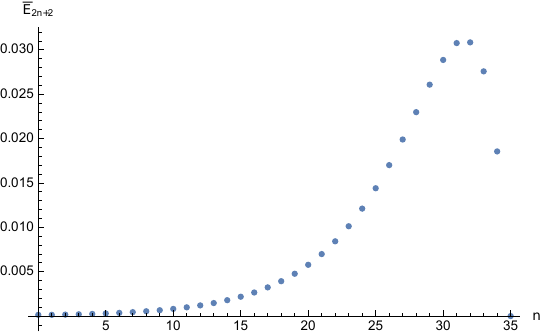}
\caption{Left: 
It is plotted the energy of odd particles $E_{2n+1}$ as a function of
the number $n$ of the decay.  Right: It is plotted the energy of even
particles $ E_{2n+2}$ as a function of the number $n$ of the decay.
Both plots are for a recursive Penrose process occurring in a
Reissner-Nordstr\"om-AdS black hole spacetime when the index $n_{\rm
c}$ is an integer, where $n_{\rm c}$ is the number for which the
charge of the black hole reaches a zero value.
The plots are for $0 \leq n \leq n_{\rm c}$.
The quantities used in
the plots are unitless, with the initial black hole mass $M_0$ serving
as the rescaling quantity. The rescaled quantities are then $\bar
E_{2n+1}=\frac{E_{2n+1}}{M_0}$, $\bar E_{2n+2}=\frac{E_{2n+2}}{M_0}$,
$\bar Q_0=\frac{Q_0}{M_0}$, $\bar e_0=\frac{e_0}{M_0}$, $\bar
l=\frac{l}{M_0}$ and $\bar r_{\rm i}= \frac{r_{\rm i}}{M_0}$. The
values used are $\bar{Q_0}=0.78$, $\bar{e_0}=0.00068$, $\bar m_0 =
0.0001$, $\bar l = 15.3$, $\bar r_{\rm i}= 5.9899$, $\alpha_2 = 0.3$,
and $\beta_2 = 1.223$.  For these values one has that the value of $n$
when the black hole electric charge is zero is $n_{\rm c} = 35$, for
which the process naturally comes to a halt.}
\label{fig:Eodd}
\end{figure}

As shown in the previous sections, the recursive process only occurs
up to $n=n_{\rm c}$. For $n\leq n_{\rm c}$, the energy of odd
and even particles,
$E_{2n+1}$ and $E_{2n+2}$, respectively,
can be obtained solving iteratively 
Eqs.~\eqref{eq:energy1_case2_penrose}
and
\eqref{eq:energy2_case2_penrose},
with the black hole charge being given by
Eq.~\eqref{eq:charge_bh_general}.  The
energies $E_{2n+1}$ and $E_{2n+2}$
are plotted in
Fig.~\ref{fig:Eodd} for specific values of the black hole's initial
mass and charge, decaying particle's initial mass and charge,
cosmological length, and parameters $\alpha_2$ and $\beta_2$. It can
be seen that the energy of odd particles starts decreasing with $n$,
becoming negative and reaching a minimum value. After this minimum,
the energy increases to zero at $n=n_{\rm c}$.
For even particles, the energy
increases with $n$ until reaching a maximum value, after which
it significantly decreases until $n=n_{\rm c}$.

Since the decay chain ends at $n=n_{\rm c}$, the energy extracted with this
process is the energy of
 the
last even particle emitted outward, i.e., the energy of particle
$2n_{\rm c} +2$.
To be precise the energy extracted is $E_{2n_{\rm c}+2}-E_0$.
From
Eq.~\eqref{eq:E_odd_penrose}, since $Q_{n_{\rm c}} = 0$, the energy
in particle 
$2n_{\rm c} +2$
is $E_{2n_{\rm c}+2} = m_{2n_{\rm c}+2}\, \sqrt{f_{n_{\rm c}}\left(r_{\rm
i}\right)}$. From Eqs.~\eqref{eq:f_RN_AdS_n},
\eqref{eq:E_odd_penrose}, and \eqref{eq:mass1_case1_penrose}, this
energy after $n_{\rm c} +1$ decays can be written as
\begin{equation}
  E_{2n_{\rm c}+2} = m_0 \,\alpha_2^{n_{\rm c}+1} \sqrt{1-2\,
  \frac{M_0+\sum_{k=0}^{n_{\rm c}-1} E_{2k+1}}{r_{\rm i}}+
  \frac{r_{\rm i}^2}{l^2}}\,.
\label{eq:energy_extracted_scenario1}
\end{equation}
This quantity is always finite and in general nonzero. However, since
$\alpha_2 < 1$,
the presence of the factor
$\alpha_2^{n_{\rm c}+1}$
means
that
the energy extracted decreases geometrically with
$n_{\rm c}$, being approximately zero for $n_{\rm c}$ large enough,
as can be seen
from Fig.~\ref{fig:Eodd}. Therefore,
to maximize the energy extracted, $n_{\rm c}$ must
assume low values, which occurs for large values of $\beta_2$ and low
ratios $\frac{Q_0}{e_0}$, see Eq.~\eqref{eq:n_critical}.
Furthermore, from Eq.~\eqref{eq:energy_extracted_scenario1}
it can be seen that if one wants a
positive net energy extraction, i.e.,  $E_{2n_{\rm c}+2}>E_0$
then one must have
the values of $n_{\rm c}$ satisfying $\alpha_2^{2n_{\rm c}}
\left(1-2\,\frac{M_0+\sum_{k=0}^{n_{\rm c}-1} E_{2k+1}}{r_{\rm i}}+
\frac{r_{\rm i}^2}{l^2}\right) > \frac{E_0^2}{\alpha_2^2 \, m_0^2}$.
In this case, there is a finite energy gain confined
to a finite space volume $V\left(r_{\rm o}\right)$, see
\cite{paper1} for the complete calculation of the
volume. Therefore, one has a black hole energy factory in this case
of
integer value of the index $n_{\rm c}$. In turn, if $\alpha_2^{2n_{\rm c}}
\left(1-2\,\frac{M_0+\sum_{k=0}^{n_{\rm c}-1} E_{2k+1}}{r_{\rm i}}+
\frac{r_{\rm i}^2}{l^2}\right) < \frac{E_0^2}{\alpha_2^2 \, m_0^2}$,
the energy $E_{2n_{\rm c}+2}$
is lower than the energy $E_0$ of the initial decaying
particle, there is no energy gain and no net energy extracted.


\subsubsection{Mass of the black hole}

\begin{figure}[h]
    \centering
    \includegraphics[width = 0.4\textwidth, height =
    0.2\textheight]{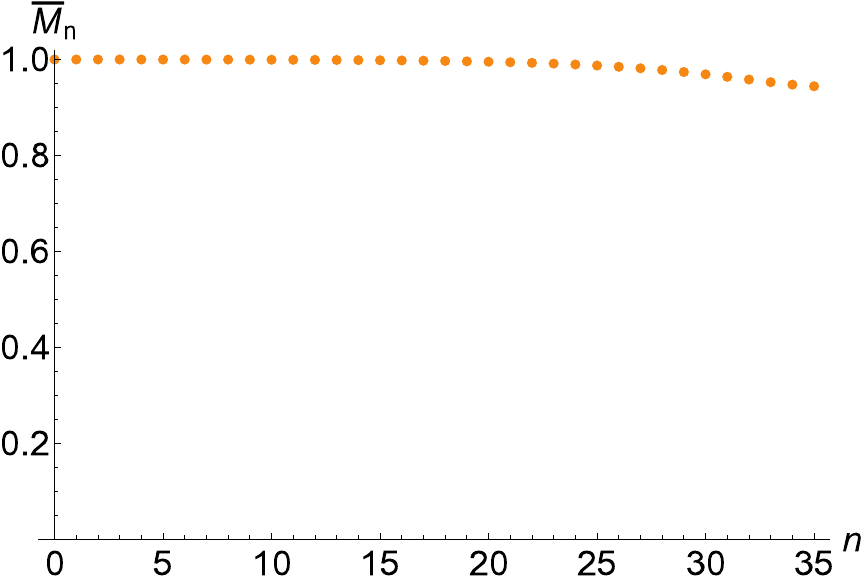}
\caption{
It is plotted the  mass of the black hole $M_n$
as a function of the number $n$ of the decay, for
a recursive Penrose process occurring in a Reissner-Nordstr\"om-AdS
black hole spacetime
when the index
$n_{\rm c}$ is an integer, where $n_{\rm c}$ is the number for which the
charge of the black hole reaches a zero value.
The plot is for $0 \leq n \leq n_{\rm c}$.
The quantities used in the plot are
unitless, with the initial black hole mass $M_0$
serving as the rescaling quantity. So, the rescaled
 quantities are
 $\bar
M_n=\frac{M_n}{M_0}$,
$\bar
E_{2n+1}=\frac{E_{2n+1}}{M_0}$, $\bar E_{2n+2}=\frac{E_{2n+2}}{M_0}$,
$\bar Q_0=\frac{Q_0}{M_0}$, $\bar e_0=\frac{e_0}{M_0}$, $\bar
l=\frac{l}{M_0}$ and $\bar r_{\rm i}= \frac{r_{\rm i}}{M_0}$. The
values used are $\bar{Q_0}=0.78$, $\bar{e_0}=0.00068$, $\bar m_0 =
0.0001$, $\bar l = 15.3$, $\bar r_{\rm i}= 5.9899$, $\alpha_2 = 0.3$,
and $\beta_2 = 1.223$. For
these values one has that the value of $n$ when the black hole
electric charge is zero is $n_{\rm c} = 35$, for which the
process naturally comes to a halt.  }
\label{fig:massBH}
\end{figure}

The mass $M$ of the black hole as a function of $n$ can also be
plotted.  This can be done knowing the energy of odd particles
corresponding to each value of $n$, see
Eq.~\eqref{eq:energy1_case2_penrose}, and using
Eq.~\eqref{eq:bh_mass2}, as it is shown in
Fig.~\ref{fig:massBH}.  It is seen
that the mass of the black hole decreases slightly along the
decay chain. The decrease in the mass
is due to the accretion of
the odd particles which themselves have negative
energy. The last particles that falls in into the black hole
is the particle $2n_{\rm c}$, but since it has negligible
relative mass it makes no impact in the final black hole mass.

\newpage

\subsubsection{Charge to mass ratio of the black hole}

The charge to mass ratio $\frac{Q_n}{M_n}$
of the black hole
at each decay $n$ is an important quantity
as it gives information of how far is the black hole from
being extremal.

\begin{figure}[h]
    \centering
    \includegraphics[width = 0.35\textwidth, height =
    0.2\textheight]{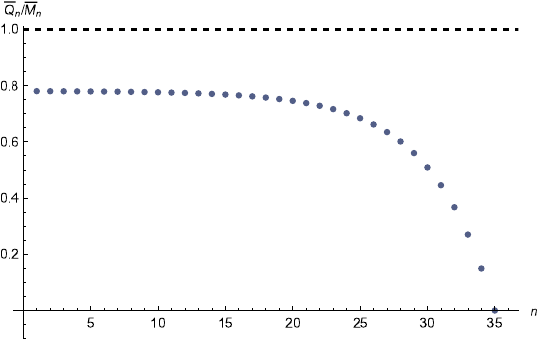}
    \hskip 2cm
    \includegraphics[width = 0.4\textwidth, height =
    0.2\textheight]{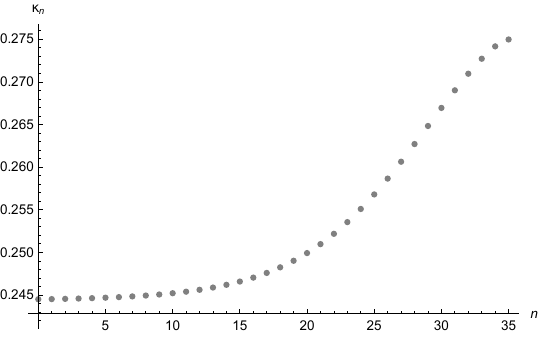}
\caption{Left: It is plotted (dark blue) the ratio between
the charge of the black hole $Q_{n}$ and the mass of the black hole $
M_{n}$, $\frac{Q_n}{M_n}$, as a function of the number $n$ of the
decay.
The ratio between $ Q_{n}$
and $M_{n}$ for the corresponding extremal black hole
is also shown (black dashed).
Right: It is plotted the surface gravity
$\kappa_n$ as a function of the number $n$ of the decay.
For an extremal black hole $\kappa=0$.
The plots are
for a recursive Penrose process occurring in a
Reissner-Nordstr\"om-AdS black hole spacetime
with $0 \leq n \leq n_{\rm c}$,
where $n_{\rm c}$ is the number for which the
charge of the black hole reaches a zero value,
here an integer.
The quantities used in the plot are
unitless, with the initial black hole mass $M_0$
serving as the rescaling quantity. So, the rescaled
 quantities are
$\bar M_n=\frac{M_n}{M_0}$, $\bar Q_n=\frac{Q_n}{M_0}$
and $\bar l = \frac{l}{M_0}$.  The
values used are $\bar{Q_0}=0.78$, $\bar{e_0}=0.00068$, $\bar m_0 =
0.0001$, $\bar l = 15.3$, $\bar r_{\rm i}= 5.9899$, $\alpha_2 = 0.3$,
and $\beta_2 = 1.223$.
For
these values one has that the value of $n$ when the black hole
electric charge is zero is $n_{\rm c} = 35$, for which the
process naturally comes to a halt.
Since we
use $\bar l = 15.3$, an
extremal black hole has $\frac{\bar Q_n}{\bar M_n} = 0.997897$
and $\kappa=0$.
}
\label{fig:ratio1}
\end{figure}

In Fig.~\ref{fig:ratio1} the relation
between the charge and
the mass of the black hole
for each $n$ is displayed. 
In the left plot of the figure
the ratio between the charge and
the mass of the black hole 
as a function of the number $n$ is plotted and seen to decrease.
This shows that the extremal limit is never obtained in  the process.
In the right plot of the figure
the
surface gravity $\kappa$ for a Reissner-Nordstr\"om black
hole, which is given
by $\kappa_n=\frac{1}{2 {r_+}_n   } \left(1+\frac{3 {r_+}_n^2}{l^2}-
\frac{Q_n^2}{ {r_+}_n^2}\right)$ at each decay $n$,
with ${r_+}_n={r_+}_n(M_n,Q_n,l)$,
is plotted 
as a function of the number $n$  and seen to increase.
Since the extremal limit has zero surface gravity, 
we see that the process never approaches an extremal
black hole.

\subsubsection{Validity of the approximation}

\begin{figure}[h]
    \centering
    \includegraphics[width = 0.4\textwidth, height =
    0.2\textheight]{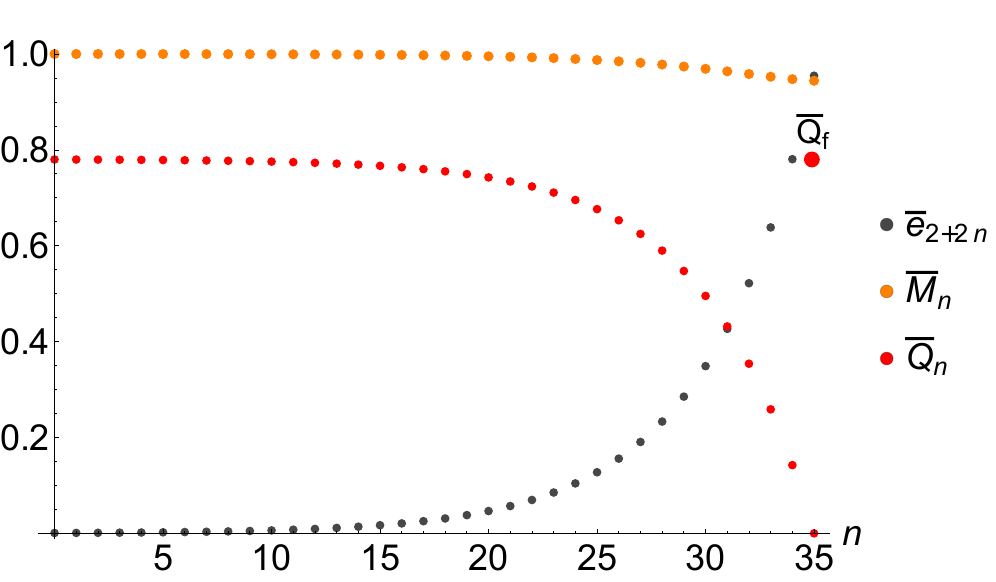}
\caption{
It is plotted the charge of even particles $ e_{2n+2}$
(gray), the mass of the black hole $M_{n}$ (orange), and the charge of
the black hole $Q_{n}$ (red), as a function of the number $n$ of the
decay, for a recursive Penrose process occurring in a
Reissner-Nordstr\"om-AdS black hole spacetime
with $0 \leq n \leq n_{\rm c}$,
where  $n_{\rm c}$ is the number for which the
charge of the black hole reaches a zero value,
here an integer.  The quantities used in the plot are unitless, with the
initial black hole mass $M_0$ serving as the rescaling quantity. So,
the rescaled quantities are $\bar e_{2n+2}=\frac{e_{2n+2}}{M_0}$,
$\bar M_n=\frac{M_n}{M_0}$, $\bar Q_n=\frac{Q_n}{M_0}$, and $\bar
e_0=\frac{e_0}{M_0}$. The values used are $\bar{e_0}=0.00068$, and
$\beta_2 = 1.223$.  For these values one has that the value of $n$
when the black hole electric charge is zero is $n_{\rm c} = 35$, for which
the process naturally comes to a halt.  Since the values of the black
hole mass $M_n$ and the particle's electric charge $e_{2n+2}$
for $n$ near $n_{\rm c} = 35$ are of
the same order, the approximations used are no longer reliable.
Nevertheless, the final configuration should be a
Reissner-Nordstr\"om-AdS black hole with electric charge
given by
$Q_{\rm f} = Q_0+e_0$.
}
\label{fig:validity1}
\end{figure}

Our approximation to study backreaction effects
relies on two related hypotheses.
One is that the particles are test particles, i.e., they do not
influence the background geometry.
The other is that the background geometry and electric
fields are spherical symmetric and are preserved along the 
recursive Penrose process. 
This approximation is valid as long as the
particles involved have masses and electric charges significantly
lower than the mass and charge of the black hole.
Once these quantities become comparable, then first, the concept of
test particle no longer applies and, second there is a preferential
direction, namely, the direction of the particle's movement. This
breaks the spherical geometry of the background, the problem is to be
treated as a two-body problem, and the electric field of the particles
must be considered to obtain a more correct description.
In Fig.~\ref{fig:validity1} the validity range of our approximation
is shown clearly.

From Eqs.~\eqref{eq:mass2_case1_penrose} and
\eqref{eq:mass1_case1_penrose}, it is shown that the masses of even
and odd particles always decrease along the decay chain, being always
smaller than the mass of the black hole, provided that the mass of the
first particle decaying is sufficiently small.
However, the electric charges of odd and even particles increase
significantly along the chain, as can be seen from
Eqs.~\eqref{eq:charge2_case2_penrose} and
\eqref{eq:charge1_case2_penrose}. It can also be inferred that, since
$\beta_2 > \beta_1$, the electric charge of even particles increases
faster than that of odd particles. In Fig.~\ref{fig:validity1}, it is
shown the electric charge of even particles along the decay chain.
Near $n_{\rm c}$, the charge of the particles reaches values close
to the mass of the black hole. Moreover, for $n$ sufficiently close to
$n_{\rm c}$, the charge of the black hole becomes of the order of the
particles' electric charge, approximately after $n=30$.
Therefore, for  $n$ near  $n_{\rm c}$ the approximation we
are using does not yield good quantitative results.
From Fig.~\ref{fig:validity1} we deduce that for $n=25$
the ratio $\frac{e_{2n+2}}{M_0}$, which
quantifies the deviation from the test-particle and spherical-symmetry
assumptions,
is about 20\% and increases
to about 100\% when $n=n_{\rm c}=35$.
This means that for 
$n$ near $n_{\rm c}$
the results are not quantitatively reliable.
For such $n$ the initial spherically symmetric
geometry becomes distorted and one should take this
into consideration.
Nevertheless, there is nothing to prevent the
decaying process and absorption of the particles by the
black hole at these values of $n$ and one should
end up, after emission of some gravitational radiation,
all the same with a spherically symmetric
configuration, i.e., a
Reissner-Nordstr\"om-AdS black hole with electric charge
given by Eq.~\eqref{eq:charge_final}, i.e., 
$Q_{\rm f} = Q_0+e_0$.

\newpage


\subsection{Case 2: Noninteger value of the index $n_{\rm c}$}

\subsubsection{Preamble}

Now, we treat the case of a noninteger value of the index $n_{\rm c}$.
The index $n_{\rm c}$ defined in Eq.~\eqref{eq:n_critical}
is the number at which the charge
$Q$ of the black hole
reaches a zero value, namely,
$n_{\rm c} = \frac{\ln{\left(1+\frac{Q_0}{e_0}\right)}}{\ln{\beta_2}}$.
Since now $n_{\rm c}$ is noninteger,
and the number of decays $n$ is an integer, 
the charge
$Q$ of the black hole is never zero and the Penrose process could
be thought to evolve for arbitrarily large $n$. 
But as we will see if that were the case,
it would lead to a gross violation
of the cosmic censorship hypothesis, 
by turning first the black hole into a naked singularity 
with mass lower than charge, followed by turning
this positive mass naked singularity into a
negative mass one. So the process has to stop
itself
just before cosmic censorship is violated.
Let us see this.

For a noninteger value of $n_{\rm c}$ the black hole electric charge changes
sign if one takes $n$ to large values.  One has that for $n_{\rm
c}^-$, defined as the greatest integer less than $n_{\rm c}$, the black hole
charge is still positive, $Q_{n_{\rm c}^-} > 0$. But, for $n_{\rm
c}^+$, defined as the lowest integer greater than $n_{\rm c}$, one finds
that the black hole charge turns negative, $Q_{n_{\rm c}^+} < 0$.
However, for $n=n_{\rm c}^-$, the electric charge of the particle
labeled $2n_{\rm c}^-+2$, which would fall into the black hole,
becomes extremely large, of the same order as the black hole mass. In
this regime the particle can no longer be treated as a test particle,
and its charge induces a substantial modification of the spacetime
geometry. The system therefore transitions from a one-body problem
with spherical symmetry to a two-body
problem. Consequently, our approximation scheme ceases to be valid,
and the calculations must be terminated at this point.

Indeed, if one would continue the calculations past $n_{\rm c}^-$
one would get
nonsensical results.
Since
for the next decay after $n_{\rm c}^-$, i.e., for
$n_{\rm c}^+$, the black hole charge turns
negative, $Q_{n_{\rm c}^+} < 0$, one finds that
the next negatively charged odd particle no longer falls
into the black hole, it suffers a repulsion and moves outward.  In
turn, the next positively charged even particle is attracted and falls
into the hole.  This state of attraction of positively even particles
would continue
until enough even particles have fallen in to reverse the
black hole charge back into a positive value, and this oscillation of
the electric charge of the black hole would go
on ad eternum.  But now
note that the black hole electric charge oscillates with increasing
amplitude for $n>n_{\rm c}$. For $n=n_{\rm c}^+ +1$, the first even
particle falls into the black hole.  Since $\beta_2 > 1$
and $\beta_1<-1$,
$e_{2n+2}$ is proportional to $\beta_2^{n+1}$, and $e_{2n+1}$ is
proportional to $\beta_1 \beta_2^{n}$, see
Eqs.~\eqref{eq:charge2_case2_penrose} and
\eqref{eq:charge1_case2_penrose}, this even particle has a very large
positive charge, much larger than the previous odd particle,
increasing by a great amount the black hole electric charge itself. On
the other hand, since there is an electric ergosphere at the decaying
position, the energy of this even particle is negative, and in
absorbing it, the black hole mass decreases.
So, as the process approaches $n=n_{\rm c}$, the black hole mass decreases
and its electric charge grows significantly. If one were to continue
the naive iteration beyond this point, the black hole would appear to
evolve into a configuration with $M<Q$, effectively becoming a naked
singularity. Such an outcome would be a blatant violation of cosmic
censorship. However, cosmic censorship is expected to hold for this
class of processes, and the apparent violation is simply an artifact
of pushing our approximations beyond their domain of validity.
Indeed, the test-particle approximation ceases to be reliable already
before reaching $n_{\rm c}^-$. Near $n_{\rm c}$, the decay products acquire
large electric charges so large that they can no longer be treated as
negligible perturbations of the background geometry. Spherical
symmetry is lost, and the problem becomes a genuine two-body
interaction in general relativity rather than a one-body system with
small test particles orbiting a fixed background. Accounting properly
for this transition naturally halts the process.
At this stage, the ratio $\frac{e_{2n+1}}{M_0}$, which
quantifies the deviation from the test-particle and spherical-symmetry
assumptions, increases with $n$, reaching values of order unity,
though still below one, as $n$ approaches $n_{\rm c}^-$.
This is the stage at which
the approximation breaks down and is no longer reliable.
One may speculate about what occurs once the two-body regime is
entered. A fully dynamical evolution admits
then the possibility that 
the
charged particle and the black hole would move
apart.
Given that the electric repulsion dominates in this regime, the
outcome should be that the two objects separate and do not
coalesce.

\subsubsection{Electric charge of the particles and
electric charge of the black hole}

\begin{figure}[h]
\centering
    \includegraphics[width = 0.4\textwidth, height =
    0.2\textheight]{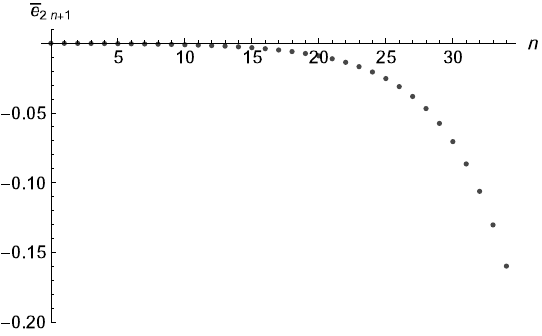}
    \hskip 2cm
    \includegraphics[width = 0.4\textwidth, height =
    0.2\textheight]{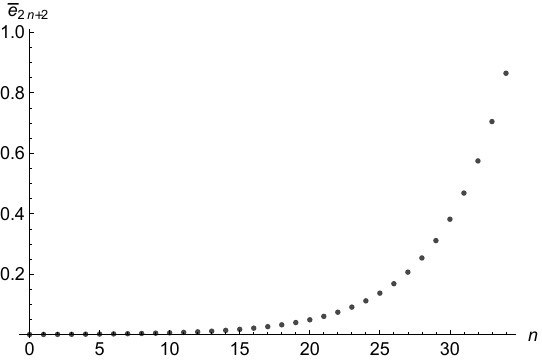}
\caption{Left: It is plotted the electric charge $e_{2n+1}$ of the
odd particles as a function of the number $n$ of the decay. Right: It
is plotted the electric charge of the even particles as a function of
the number $n$ of the decay.
The plots are 
for a recursive Penrose process occurring in a
Reissner-Nordstr\"om-AdS black hole spacetime
for $0 \leq n \leq n_{\rm c}^-$,
where $n_{\rm c}^-$ is the greatest
integer lower than $n_{\rm c}$, with
$n_{\rm c}$ being the number for which the
charge of the black hole reaches a zero value,
here not an integer. The quantities used
are unitless, with the initial black hole mass $M_0$ serving as the
rescaling quantity.  So the rescaled electric charges $\bar
e_{2n+1}=\frac{e_{2n+1}}{M_0}$ and $\bar
e_{2n+2}=\frac{e_{2n+2}}{M_0}$ are the quantities plotted as a
function of $n$. The other rescaled quantities are $\bar
Q_0=\frac{Q_0}{M}$ and $\bar e_0=\frac{e_0}{M_0}$.  The values used
are $\bar{Q_0}=0.78$, $\bar{e_0}=0.00068$, and $\beta_2 = 1.227$.  For
these values one has that the value of the index $n_{\rm c}$ is
$n_{\rm c} = 34.5$ and so the process stops at $n=n_{\rm c}^-=34$.}
\label{fig:chargeofparticlesnoninteger}
\end{figure}

To study the evolution of the particles electric
charge we again use iteratively
Eqs.~\eqref{eq:charge2_case2_penrose} and
\eqref{eq:charge1_case2_penrose}.  We see that odd particles have a
negative electric charge, whose absolute value increases geometrically
with the number $n$, while even particles have a positive electric
charge, increasing also geometrically. The electric charge of the
particles and the electric charge of the black hole are coupled
through Eq.~\eqref{eq:charge_bh_general}, therefore one must compute
the different charges consistently.  In
Fig.~\ref{fig:chargeofparticlesnoninteger}, it is shown the evolution of
electric charge of the particles $e_{2n+1}$ and $e_{2n+2}$ as a
function of the number of decays $n$.

\newpage

\begin{figure}[h]
    \centering
    \includegraphics[width = 0.42\textwidth, height =
    0.2\textheight]{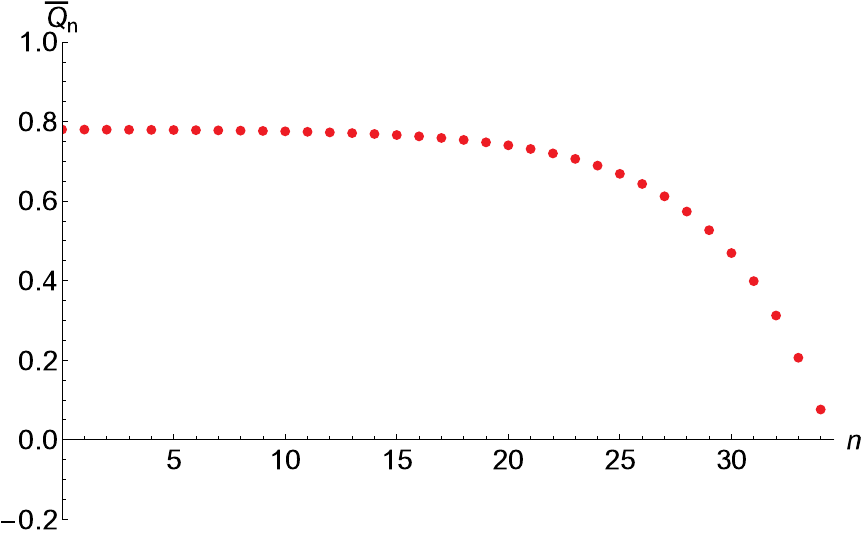}
\caption{It is plotted the
charge of the black hole $
Q_n$, as a function of the number $n$ of the decay, for a recursive
Penrose process.
The plot is
for a recursive Penrose process occurring in a
Reissner-Nordstr\"om-AdS black hole spacetime
for $0 \leq n \leq n_{\rm c}^-$,
where $n_{\rm c}^-$ is the greatest
integer lower than $n_{\rm c}$, with
$n_{\rm c}$ being the number for which the
charge of the black hole reaches a zero value,
here not an integer.
The quantities used in the plots are unitless, with the initial
black hole mass $M_0$
serving as the rescaling quantity.
So, the rescaled quantities are $\bar Q_n=\frac{Q_n}{M_0}$,
$\bar Q_0=\frac{Q_0}{M_0}$, and $\bar e_0=\frac{e_0}{M_0}$. The values
used are $\bar{Q_0}=0.78$, $\bar{e_0}=0.00068$, and $\beta_2 = 1.227$.
For
these values one has that the value of the index $n_{\rm c}$ is $n_{\rm c} =
34.5$ and so the process stops at $n=n_{\rm c}^-=34$.
}
\label{fig:charge3}
\end{figure}

To study the evolution of the black hole electric charge we use
 Eq.~\eqref{eq:charge_bh_general}
 iteratively. In
Fig.~\ref{fig:charge3}
the black hole charge itself $Q_n$ is shown as
a function of the number $n$
of the decay up to $n_{\rm c}^-$. 
At $n_{\rm c}^-$ note now that
the Penrose process stops here,
in principle no more particles
fall in in which case the black hole
acquires here its final mass.
What happens next, although our calculation cannot
follow it, is that the two objects move apart
due to the electric repulsion. The charge of the black
hole is about zero and the charge of the particle is about $Q_0$,
so that we are able to write
\begin{equation}
Q_{\rm f} = q,\quad\quad\quad e_{2n_{\rm c}^-}=Q_0- (q-e_0)\,,
   \label{eq:charge_final1}
\end{equation}
where $q$ is a charge satisfying $q\ll Q_0$ but can be somewhat
larger than $e_0$. 
Notice that charge conservation must still hold, meaning
that $Q_{\rm f} + e_{2n_{\rm c}^-}= Q_0 + e_0$.

\newpage

\subsubsection{Analysis of the turning points}

In this case of a noninteger value of the index $n_{\rm c}$,
the electric charge of the black hole
remains always nonzero. Therefore,  an inner turning
point $r_{\rm i}$ always exists, solution of
Eq.~\eqref{eq:turning_points_penrose},
where the decays occur, and an outer turning point $r_{\rm o}$ also
always exists
corresponding to the largest solution of
Eq.~\eqref{eq:turning_points_penrose}. Both turning points are finite
throughout the decay chain, with the position of $r_{\rm i}$
being
fixed, and the position
of $r_{\rm o}$ changing, depending
on the mass and electric charge of the outgoing particles.  An
explicit expression for the position of the outer turning point is
derived in \cite{paper1}.
The turning point $r_{\rm o}$
can in general oscillate between low
and large values of $r$ depending on the parameters of the black hole
and the particles. For sufficiently large values of
energy per unit mass of the particles, the position of this outer
turning point increases. However, since
our approximation is valid up to an $n$ approaching
$n_{\rm c}^-$,
and after which
cosmic censorship is violated, the process has a finite
number of steps, one has that the position of the outermost
turning point  $r_{\rm o}$
is finite, ensuring particle confinement in a finite
volume.

\subsubsection{Energy extraction}

\begin{figure}[h]
    \centering
\includegraphics[width = 0.42\textwidth, height = 0.2
\textheight]{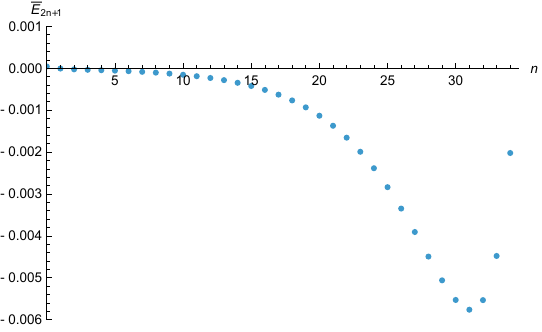}
\includegraphics[width = 0.42\textwidth, height = 0.2
\textheight]{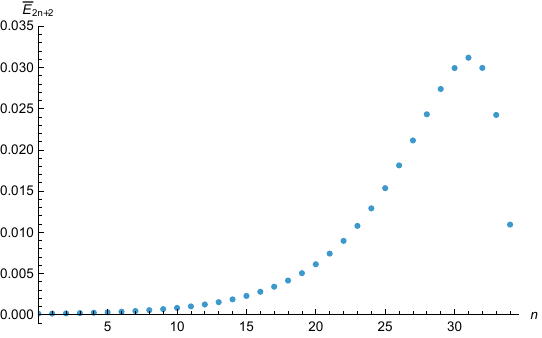}
\caption{
Left:
It is plotted the energy of odd particles $E_{2n+1}$ as a function of
the number $n$ of the decay.  Right: It is plotted the energy of even
particles $ E_{2n+2}$ as a function of the number $n$ of the decay.
The plots are 
for a recursive Penrose process occurring in a
Reissner-Nordstr\"om-AdS black hole spacetime
for $0 \leq n \leq n_{\rm c}^-$,
where $n_{\rm c}^-$ is the greatest
integer lower than $n_{\rm c}$, with
$n_{\rm c}$ being the number for which the
charge of the black hole reaches a zero value,
here not an integer.
The quantities used in the plots are
unitless, with the initial black hole mass $M_0$ serving as the
rescaling quantity.  So, the rescaled quantities are $\bar
E_{2n+1}=\frac{E_{2n+1}}{M_0}$, $\bar E_{2n+2}=\frac{E_{2n+2}}{M_0}$,
$\bar Q_0=\frac{Q_0}{M_0}$, $\bar e_0=\frac{e_0}{M_0}$, $\bar
l=\frac{l}{M_0}$ and $\bar r_{\rm i}= \frac{r_{\rm i}}{M_0}$. The
values used are $\bar{Q_0}=0.78$, $\bar{e_0}=0.00068$, $\bar m_0 =
0.0001$, $\bar l = 15.3$, $\bar r_{\rm i}= 5.9899$, $\alpha_2 = 0.3$,
and $\beta_2 = 1.227$.
For
these values one has that the value of the index $n_{\rm c}$ is $n_{\rm c} =
34.5$ and so the process stops at $n=n_{\rm c}^-=34$.
}
\label{fig:E_non_integer}
\end{figure}

For $n\leq n_{\rm c}^-$, the energy of odd and even particles is given by
Eqs.~\eqref{eq:energy1_case2_penrose} and
\eqref{eq:energy2_case2_penrose}, respectively. These energies are
plotted in Fig.~\ref{fig:E_non_integer} for specific
values of the black hole's initial mass $M_0$
and charge $Q_0$, decaying
particle's initial mass $m_0$
and charge $e_0$, cosmological length
$l$, and decay
parameters $\alpha_2$ and $\beta_2$.
The energy $E_{2n+1}$ of odd
particles decreases with $n$, becomes negative and reaches a
minimum value. 
The energy $E_{2n+2}$ of the outgoing particles, i.e., the
ones carrying the energy extracted,
is positive, since
$e_{2n+2}\, Q_n > 0$ when the outgoing
particle is even, as mentioned previously. 

\newpage

\subsubsection{Mass of the black hole}

\begin{figure}[h]
    \centering
    \includegraphics[width = 0.38\textwidth, height =
    0.2\textheight]{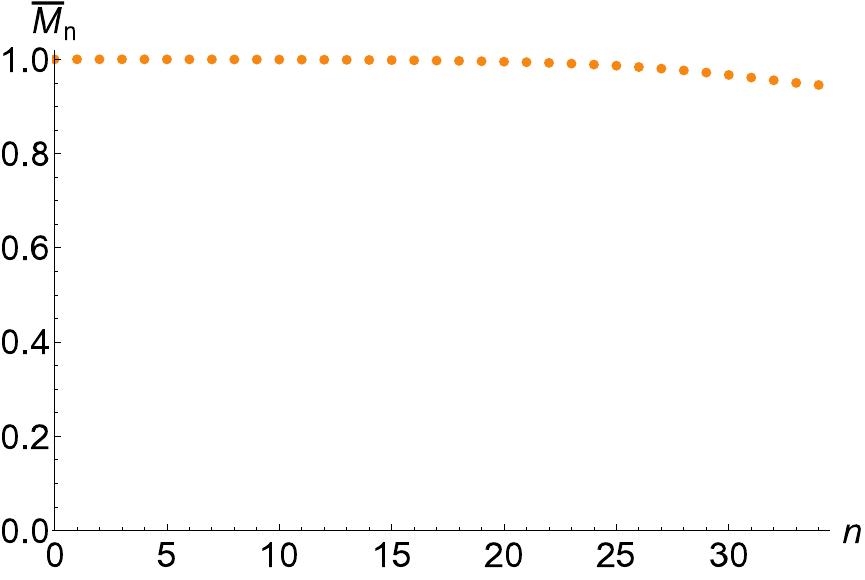}
\caption{
It is plotted the mass of the black hole $M_n$ as a function of the
decay number $n$.
The plot is
for a recursive Penrose process occurring in a
Reissner-Nordstr\"om-AdS black hole spacetime
for $0 \leq n \leq n_{\rm c}^-$,
where $n_{\rm c}^-$ is the greatest
integer lower than $n_{\rm c}$, with
$n_{\rm c}$ being the number for which the
charge of the black hole reaches a zero value,
here not an integer.
The quantities used in the plot are
unitless, with the initial black hole mass $M_0$
serving as the rescaling quantity.
So, the rescaled quantities are
$\bar M_n=\frac{M_n}{M_0}$,
$\bar
E_{2n+1}=\frac{E_{2n+1}}{M_0}$, $\bar E_{2n+2}=\frac{E_{2n+2}}{M_0}$,
$\bar Q_0=\frac{Q_0}{M_0}$, $\bar e_0=\frac{e_0}{M_0}$, $\bar
l=\frac{l}{M_0}$ and $\bar r_{\rm i}= \frac{r_{\rm i}}{M_0}$. The
values used are $\bar{Q_0}=0.78$, $\bar{e_0}=0.00068$, $\bar m_0 =
0.0001$, $\bar l = 15.3$, $\bar r_{\rm i}= 5.9899$, $\alpha_2 = 0.3$,
and $\beta_2 = 1.227$.
For
these values one has that the value of the index $n_{\rm c}$ is $n_{\rm c} =
34.5$ and so the process stops at $n=n_{\rm c}^-=34$. }
\label{fig:massBH2}
\end{figure}

The mass of the black hole as a function of $n$
is computed iteratively, using 
Eq.~\eqref{eq:bh_mass2}
and Eq.~\eqref{eq:energy1_case2_penrose}, as it is shown in
Fig.~\ref{fig:massBH2}.  
It can be
seen that the mass of the black hole decreases along the decay chain,
due to the accretion of particles with negative energy.

\subsubsection{Charge to mass ratio of the black hole}

\begin{figure}[h]
    \centering
    \includegraphics[width = 0.35\textwidth, height =
    0.2\textheight]{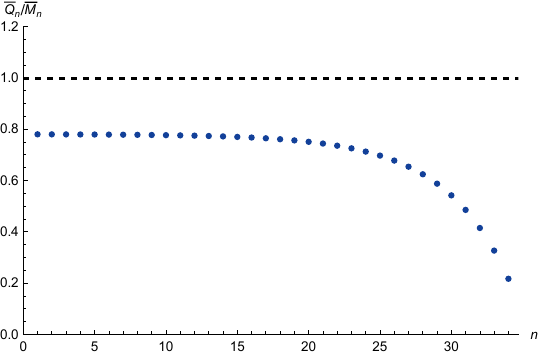}
\hskip 1.5cm
\centering
    \includegraphics[width = 0.4\textwidth, height =
    0.2\textheight]{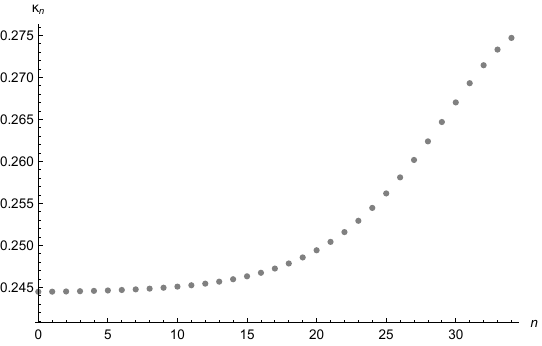}
\caption{
Left:
It is plotted (dark blue)  the ratio between the 
charge of the black hole $Q_{n}$ and the mass of the
black hole $M_{n}$, as a function of the number $n$ of the
decay. The ratio between $ Q_{n}$
and $M_{n}$ for the corresponding extremal black hole
is also shown (black dashed).
Points below it are black holes,
above it would be naked singularities with charge greater
than mass.
Right: It is
plotted the surface gravity $\kappa_n$
as a function of the number $n$ of the decay.
For an extremal black hole $\kappa=0$.
The plots are 
for a recursive Penrose process occurring in a
Reissner-Nordstr\"om-AdS black hole spacetime
for $0 \leq n \leq n_{\rm c}^-$,
where $n_{\rm c}^-$ is the greatest
integer lower than $n_{\rm c}$, with
$n_{\rm c}$ being the number for which the
charge of the black hole reaches a zero value,
here not an integer.
The quantities used in the plots are
unitless, with the initial black hole mass $M_0$
serving as the rescaling quantity.
So, the rescaled quantities are 
$\bar M_n=\frac{M_n}{M_0}$, $\bar
Q_n=\frac{Q_n}{M_0}$ and $\bar l = \frac{l}{M_0}$,
$\bar \kappa=\frac{\kappa}{M_0}$.
The values used are $\bar{e_0}=0.00068$, and
$\beta_2 = 1.227$
The ratio between $\bar Q_{n}$ and $\bar M_{n}$ that leads to an
extremal black hole here is $\frac{\bar Q}{\bar M} = 0.997897$, for
$\bar l = 15.3$ that we use, see the black dashed line.  For these
values one has that the value of the index $n_{\rm c}$ is $n_{\rm c} =
34.5$ and so the process stops at $n=n_{\rm c}^-=34$.
}
\label{fig:ratio2}
\end{figure}

The charge to mass ratio $\frac{Q_n}{M_n}$
of the black hole
at each decay $n$ is an important quantity
as it gives information of how far is the black hole from
being extremal.
In Fig.~\ref{fig:ratio2} the relation
between the charge and
the mass of the black hole
for each $n$ is displayed. 
In the left plot of the figure
the ratio between the charge and
the mass of the black hole 
as a function of the number $n$ is shown and seen to decrease.
This shows that the extremal limit is never obtained in  the process.
So, the black hole
does not turn into a naked singularity
with charge greater than mass.
The dashed black line is the extremal black hole
line, line below which the configurations are
black holes and above which the configurations
would be naked singularities.
The dark blue line
shows the charge to mass
ratio  along the decay chain.
For $0<n \leq n_{\rm c}^-$
the ratio is always below the line.
In the right plot of the figure
the
surface gravity $\kappa_n$ for a Reissner-Nordstr\"om black
hole, which is given
by $\kappa_n=\frac{1}{2 {r_+}_n   } \left(1+\frac{3 {r_+}_n^2}{l^2}-
\frac{Q_n^2}{ {r_+}_n^2}\right)$ at each decay $n$,
with ${r_+}_n={r_+}_n(M_n,Q_n,l)$,
is plotted 
as a function of the number $n$  and seen to increase.
For values of $n$ for
which our approximation holds, the surface gravity
remains
nonzero, which is consistent with the conclusion that
extremality is not approached for $n_{\rm c}^-$.

\newpage

\subsubsection{Validity of the approximation}

Our backreaction approach takes into account the effects of the
particles energies and electric charges on the black hole mass and
electric charge.  The approach is valid as long as the energies and
charges of the particles are much lower than the mass and charge of
the black hole.  In the later stages of the absorption of the
particles this starts to be invalid, the test
character of the particles is lost, i.e., the particles themselves
start to be source of gravitational and electric field, and
consequently the system looses spherical symmetry which was assumed
from the start.

\begin{figure}[h]
    \centering
\includegraphics[width = 0.45\textwidth, height =
0.2\textheight]{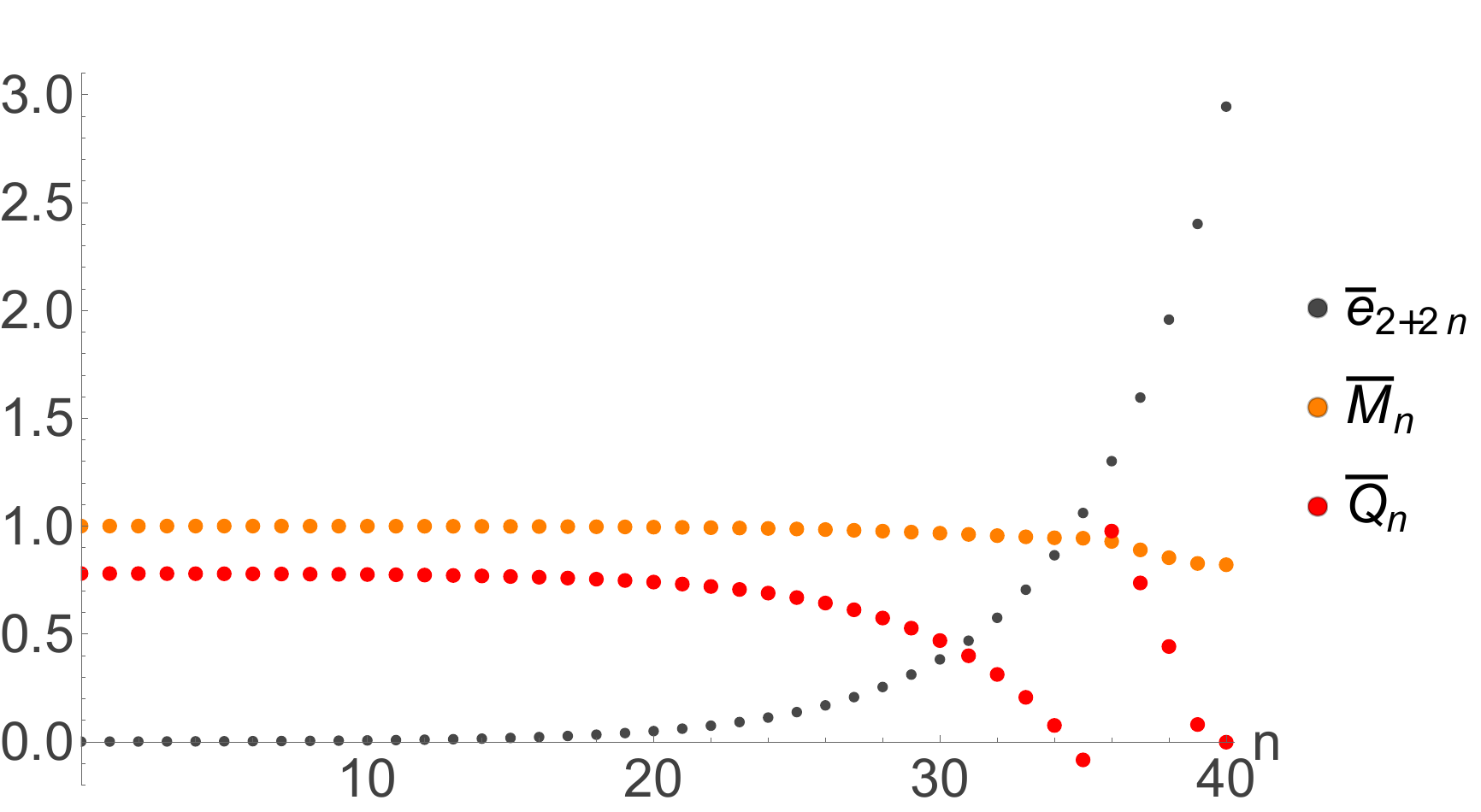}
\caption{
It is plotted the
charge of even particles 
$
e_{2n+2}$  (gray),
the  mass of the black hole $M_{n}$
(orange), and the  charge of the black hole $Q_{n}$
(red), as a function of the number $n$ of the decay
The plots are 
for a recursive
Penrose process occurring in a Reissner-Nordstr\"om-AdS black hole
spacetime.
The plots extend beyond $n_{\rm c}^-$,
where $n_{\rm c}^-$ is the greatest
integer lower than $n_{\rm c}$, with
$n_{\rm c}$ being the number for which the
charge of the black hole reaches a zero value,
here not an integer.
In this way the validity of the approximation is displayed clearly. 
The quantities used in the plots are
unitless, with the initial black hole mass $M_0$
serving as the rescaling quantity.
So, the rescaled quantities are $\bar e_{2n+2}=\frac{e_{2n+2}}{M_0}$,
$\bar M_n=\frac{M_n}{M_0}$, $\bar Q_n=\frac{Q_n}{M_0}$, and $\bar
e_0=\frac{e_0}{M_0}$. The values used are $\bar{e_0}=0.00068$, and
$\beta_2 = 1.227$.
For these values one has that the value of the index $n_{\rm c}$ is
$n_{\rm c} = 34.5$ and so the process stops at $n=n_{\rm c}^-=34$.  To
understand the validity of the approximation we make the plots past
$n=n_{\rm c}^-=34$.  For $n>n_{\rm c}^+=35$ a naked singularity with
charge greater than mass appears, then for $n>42$ a naked singularity
with negative mass appears. Due to the approximations, the
calculations must be stopped at $n=n_{\rm c}^-=34$ and so these naked
singularities never appear in reality and cosmic censorship holds.
}
\label{fig:validity2}
\end{figure}

To give more details,
in Fig.~\ref{fig:validity2} we plot 
the
charge of even particles 
$e_{2n+2}$,
the  mass  $M_{n}$ of the black hole, and the
charge $Q_{n}$ of the black
hole, all in units of
the initial black hole mass $M_0$,
as a function of the decay number up to large $n$.
The
charge of odd particles $e_{2n+1}$
is not plotted because, 
although it increases along the
chain, nevertheless,
since $\beta_2 > \beta_1$, the
electric charge of even particles increases faster,
see Eqs.~\eqref{eq:charge2_case2_penrose} and
\eqref{eq:charge1_case2_penrose}. Thus, for
the purposes here it is enough to plot 
$e_{2n+2}$.
It can be seen that close to
$n=n_{\rm c}^-$, the charge $e_{2n+2}$
of the particles reaches values close to the
mass $M_{n}$ of the black hole. Moreover, for
some $n<n_{\rm c}^-$ the electric charge $Q_{n}$
of the black hole 
becomes of the order of the particles' electric charge,
and at $n=n_{\rm c}^-$ the electric charge $Q_{n_{\rm c}^-}$
is negligibly small.
At this point, the gravitational and electric fields of
the particles must be
taken into account
in order to obtain a more precise description,
as the particles are not anymore test particles, and the spherical
symmetry of the system is broken.
Thus the calculations must be stopped at most at
$n=n_{\rm c}^-$. If one would continue the
calculations
we see from the figure
that for $n>n_{\rm c}^+$ a naked singularity with charge greater than mass
would be formed, and for even higher $n$ a
negative mass singularity would appear. This would not make sense
as cosmic censorship must hold in
this process. Fortunately, the approximations
dictate their own demise, indeed for
$n=n_{\rm c}^-$ the problem is a two-body problem, and
the two body configuration evolves somehow into
another configuration which surely
is not a naked singularity.

We can here make a quantitative estimate
displaying the breakdown of the approximation. 
We see from
Fig.~\ref{fig:validity2} that the ratio $\frac{e_{2n+2}}{M_0}$, which
quantifies the deviation from the test-particle and spherical-symmetry
assumptions, increases with $n$. This ratio is 
$\frac{e_{2n+2}}{M_0}= 5\%$
for $n=20$, 
$\frac{e_{2n+2}}{M_0}=40 \%$
for $n=30$, and 
$\frac{e_{2n+2}}{M_0}= 90\%$
for $n=n_{\rm c}^-=34$.
Qualitatively one can trust the results
for $n=n_{\rm c}^-$ but quantitatively
the error is large,  from then on the approximation is no more reliable,
the process evolves into a clear two-body problem
which requires other calculational means.
It is plausible
that the
charged particle and the black hole move apart
given that the
electric repulsion dominates in this regime.

\newpage
\section{Conclusions}
\label{sec:concl}

\vskip -0.15cm

We have analyzed the recursive Penrose process for electrically
charged particles in a Reissner-Nordstr\"om-AdS black hole spacetime,
taking into account backreaction of the Penrose process on the
spacetime metric, both on the mass of the black hole and on its
electric charge. This analysis was performed considering that the
confinement is due to the negative cosmological constant of the
asymptotically AdS spacetime.

Defining the index $n_{\rm c}$ as the value for which the black hole's
charge is zero, $Q_{n_{\rm c}}=0$, we have seen that there are two
separate cases.  Case 1 is when the index $n_{\rm c}$ is an integer.
Due to the backreaction effects, the mass and electric charge of the
black hole decrease along the decay chain and the recursive Penrose
process must stop after $n_{\rm c} +1$ decays.  The result of this
process is, in general, an emitted particle with finite energy, which
is then reflected inwards due to the outermost turning point. This
particle then falls into the black hole, since there is no inner
turning point after the black hole becomes uncharged. After falling
into the black hole, we end with a charged black hole, with charge
$Q_0 +e_0$, i.e., the charge of the initial black hole plus the charge
of the original decaying particle that started the recursive
process.
Thus,
in this case 1 the energy extracted is finite, and so we can talk of
an energy factory, not of a black hole bomb, as was found in a
previous paper not taking onto account backreaction effects.
Within this case, 
we have also found a fine-tuned possibility in which
the
even particle $2 n_{\rm c} +2$ cannot be emitted outwards,
but instead stays at rest at the decaying position, while the
particle $2 n_{\rm c} +1$ falls into the black hole, the
final configuration being 
and uncharged black hole with a particle with charge $Q_0 + e_0$ at
rest in the decay position, i.e., a black hole with small hair. 
However this configuration is unstable.
Case 2 is when the index $n_{\rm c}$ is noninteger.  We have found
that the recursive Penrose process must stop after $n_{\rm c}^-$
decays, where $n_{\rm c}^-$ is the greatest
integer lower than $n_{\rm c}$,
since after that the charge of the particles involved is
sufficiently high to turn the problem into a two-body problem and the
approximations that were assumed from the start are not valid anymore.
Namely, the test particle and the spherical symmetry approximations
break down and no more particles fall to the black hole.  This ensures
that cosmic censorship is obeyed.  The full two-body problem in
general relativity must provide a stopping mechanism for the recursive
decay chain.  In the end of the process, one gets a black hole with
arbitrarily small positive charge and a particle with high electric
charge of about $Q_0$. Electric repulsion should make the two bodies
fall apart.  Since charge conservation still holds, one has that the
sum of the particle and the black hole's charges is $Q_0 + e_0$.

Since in both cases 1 and 2 the process must stop after a finite
number of decays, the position of the outermost turning point is
always finite. Therefore, all the results obtained are effectively
independent of having a turning point or a reflective mirror placed at
some finite radius, which was not the case without considering
backreaction effects.

As a final remark, we have traced how a proper account
of backreaction in the recursive Penrose process
makes a black hole bomb impossible.

\acknowledgments{We acknowledge financial support from Funda\c c\~ao
para a Ci\^encia e Tecnologia - FCT through the
project~No.~UIDB/\break 00099/2025.}

\appendix

\section{Reflective mirror}
\label{app:mirror}

Let us consider the situation in which a reflective mirror is placed
at a radius $r_{\rm m}$. This equally applies for cases 1 and 2
previously discussed.  Two things can happen after the decay of
particle $2n$, with $n<n_{\rm c}$: if the turning point is before the
mirror, i.e., $r_{\rm o} \left(n\right) < r_{\rm m}$, particle $2n+2$
will bounce back from the turning point, reaching $r_{\rm i}$ with
zero velocity; if the mirror is placed before the turning point, i.e.,
$r_{\rm o} \left(n\right) > r_{\rm m}$, particle $2n+2$ will bounce
back from the reflective mirror, in such a way that it reaches again
the inner turning point $r_{\rm i}$ with zero velocity. In brief,
particle confinement and, as a consequence, a recursive chain of
decays, would exist either due to the reflective mirror or due to the
outer turning point with finite position. Therefore, nothing
qualitatively changes by including a reflective mirror. This is a key
difference from the analysis performed in \cite{paper1} neglecting
backreaction. When backreaction is neglected, the recursive process
can occur up to $n \to \infty$, for which the energy of even particles
diverges, and consequently the position of the turning point also
diverges, see \cite{paper1} for the full expression of the turning
point for large energies. While this would still enable a recursive
process, the volume where particles are confined would diverge as $n
\to \infty$, while in the presence of a mirror particles remain always
in a finite volume.  In our case, backreaction imposes a finite limit
for the number of decays $n$, leading to a finite position of the
turning point throughout the whole process and implying that particles
are always confined to a finite volume, making the distinction between
the situations with a mirror and without the mirror irrelevant.


\end{document}